\documentclass[times,authoryear]{elsarticle}

\usepackage{jasr}
\usepackage{framed,multirow}

\usepackage{amssymb}
\usepackage{latexsym}

\usepackage{url}
\usepackage{xcolor}
\definecolor{newcolor}{rgb}{.8,.349,.1}

\usepackage[citebordercolor=white]{hyperref}

\usepackage{graphicx}
\usepackage{amssymb}
\usepackage{amsmath}
\usepackage[subrefformat=parens]{subcaption}
\usepackage{comment}
\usepackage{bm}

\usepackage{multirow}
\usepackage{lipsum}

\journal{Advances in Space Research}

\begin{document}

\verso{Shun Isobe \textit{etal}}

\begin{frontmatter}


\title{Formation Keeping Control for Deorbiting an Uncooperative Satellite\\ by Laser Ablation}

\author[label1]{Shun Isobe\corref{cor}}
\ead{isobe.shun.439@s.kyushu-u.ac.jp}
\author[label1]{Yasuhiro Yoshimura}
\ead{y.yoshimura.a64@m.kyushu-u.ac.jp}
\author[label1]{Toshiya Hanada}
\ead{hanada.toshiya.293@m.kyushu-u.ac.jp}
\author[label2]{Yuki Itaya}
\ead{itaya-yuki@sptvjsat.com}
\author[label2]{Tadanori Fukushima}
\ead{t-fukushima@sptvjsat.com} 
\address[label1]{Kyushu University, 744 Motooka Fukuoka, Fukuoka 819-0395, Japan}
\address[label2]{SKY Perfect JSAT Corporation, 8-1 Akasaka 1-chome, Minato-ku, Tokyo 107-0052, Japan}

\cortext[cor]{Corresponding author at:
  Kyushu University, 744 Motooka Fukuoka, Fukuoka 819-0395, Japan}

\begin{abstract} 

This paper proposes the formation keeping control law for deorbiting debris by a laser ablation. 
Laser ablation is vital technology for contactless active debris removal, where a chaser satellite with a laser system irradiates laser pulses to a target object to generate the ablation force for deorbiting.  The deorbiting force decelerates the target, and the chaser must maintain its relative position and continue irradiating.  In other words, both the chaser and the target are supposed to be deorbited simultaneously, where both have accelerations.  Although conventional formation flying missions assume that only a chaser maneuvers, the formation flying in this paper considers that both a chaser and a target have accelerations.
Thus, this paper derives the relative equations of motion between the chaser and the target in powered flight and their analytical solution using relative orbital elements.  A control law based on the analytical solution is proposed, which determines the timings and directions of the laser ablation and the electrical thrust so that the formation periodically returns to a desired formation. 
Numerical simulations first examine the control law in two cases with different maneuver timings.
Then, a Monte Carlo simulation is performed to verify the effectiveness of the control law for a variety of desired formations.

\end{abstract}


\end{frontmatter}


\section{Introduction}

%
Numerous nonfunctional objects such as defunct satellites, rocket upper stages, and fragments, remain in near-Earth orbit, which are generated from on-orbit collisions or breakups. To remediate the orbital environment, an annual active debris removal (ADR) of 5--10 objects from the low Earth orbit is required~\citep{liou2011active}.  Several ADR methods, such as electrodynamic tethers and robotic arms, have been proposed and experimented on orbit~\citep{mcknight2019handbook, liou2011engineering} including the RemoveDEBRIS mission~\citep{forshaw2016removedebris} and ELSA-d~\citep{blackerby2019elsa}.  Because these methods require direct-contact operation, they have potential collision risks.  In addition, it is quite difficult to capture debris if they are uncooperative~\citep{Sasaki2021}.  On the other hand, ADR methods that use a laser have an advantage in contactless operations~\citep{tsunoend, shirasawaconcept}.  Thus, the contactless ADR method has a lower risk of functional loss due to accidental collisions.

%
Laser ablation is the process of removing materials in the vaporized or ionized state from a solid surface by irradiating it with a high-intensity laser beam~\citep{vasile2014improved}.  Ablation of materials by laser beam with a fluence over the threshold leads to complex light-matter interactions, in which the exposed material is transformed to a jet of vapor or plasma.  The ablation force is generated as the reaction force of the plasma/gas ejected from the surface.
\citet{tsuno2020impulse} measured the ablation force by Neodymium-doped yttrium aluminum garnet (Nd:YAG) pulse laser irradiating a 7075 aluminum in vacuum.  Nd:YAG is the most promising high-power pulse laser for space applications.  The laser ablation force of $0.72~\rm{mN}$ has been demonstrated, supposing a $20~\rm{\mu Ns/J}$ irradiation energy and a laser power of $1~\rm{J}$ at the $36~\rm{Hz}$ laser.  \citet{chang2014experimental} dealt with the experimental study of ablation plume from aluminum surface and revealed that the plume growth is almost vertical direction of the target surface regardless of the laser irradiating angle.  The results imply that the ablation force to change the motion of the target satellite can be generated by selecting a proper laser irradiation point in the target surface.  In the same way, the orbit transfer without changing attitude can be realized by selecting the intersection point of the irradiation surface and perpendicular line from the center of mass.

%
Laser ablation propulsion is a propulsion concept with a 49-year history~\citep{phipps2010laser}.
\citet{phipps1996orion,phipps2012removing} proposed a ground-based laser debris removal.  The ground-based laser system irradiates high power pulsed laser to the objects on orbit, depriving the orbital energy and lowering the objects for re-enter.  A $1.06~\rm{\mu m}$, $15~\rm{kJ}$, and $10~\rm{ns}$ laser with $30~\rm{kW}$ average power is studied to reduce $1-10~\rm{cm}$ debris by 80\% in two years.
As a space-based laser application, \citet{vetrisano2016asteroid} dealt with the control of orbit and rotation by laser ablation for asteroids.  The detumbling of an asteroid in a few days has been demonstrated using the proposed guidance, navigation, and control system with a moderate size laser.

%
This study deals with the ADR method by the space-borne laser.  \citet{tsunoend} showed the conceptual study of the laser ADR method and the following three advantages: contactless debris removal, adaptability to tumbling target objects, and no need to carry deorbit fuel.  \citet{sakai2022contactless} proposed the contactless attitude control method by laser ablation.  This controller has been demonstrated assuming a $150~\rm{kg}$ class object and contributes to efficiently deorbit space debris.
For ADR using laser ablation, an chaser satellite with a laser system irradiates laser pulses to a target object to generate the ablation force for deorbiting.  The deorbiting force decelerates the target, and the chaser satellite must maintain formation with respect to the target.  In other words, this ADR method is a mission of a powered formation flying of both the target object and the chaser satellite.

%
Formation flying techniques such as reconfiguration and maintenance have been studied over the years~\citep{alfriend2009spacecraft, goodman2006history}.  Many control laws are proposed based on linearized relative motion models, such as the Hill--Clohessy--Wiltshire equation~\citep{clohessy1960terminal}.  On the other hand,~\citet{d2010autonomous} defined relative orbital elements (ROE), which can visualize the relative orbit and are valid for large distances between the satellites.  \citet{gaias2015impulsive} addressed multi-impulsive schemes by ROE for formation reconfiguration. \citet{larbi2012concept} derived the concept of nonimpulsive thrust maneuvers using ROE.  \citet{di2018continuous} derived ROE-based linearized equations of relative motion and addressed the computation of semi-analytical control solutions for formation reconfiguration using piecewise continuous thrust.  These control methods assume that only either satellite maneuvers.  However, for ADR using laser ablation, both the chaser satellite and the target are supposed to have external accelerations by low-thrust electrical propulsion and laser ablation force, respectively.

%
In this context, this paper derives the relative equations of motion between the chaser and the target in powered flight and proposes a control law for the simultaneous deorbit.  The external accelerations of the chaser and target are assumed to be low continuous electrical thrust and ablation force, respectively.  Using the Gauss variational equations (GVE), the equations of relative motion and their analytical solution are derived.  Then, this paper proposes a control law for simultaneous deorbit, which determines the timings and directions of the ablation force and the electrical thrust so that the formation periodically returns to a desired formation.  
Considering practical operation of the chaser with a laser system, laser ablation and electrical propulsion are assumed to be operated exclusively.  That is, laser ablation and electric propulsion do not operate simultaneously due to power constraints on the chaser.
Furthermore, the proposed controller compensates for control error due to uncertainty in the magnitude of laser ablation and orbital perturbations.  
The proposed control law is flexible in terms of the desired formation and maneuver timings, practically contributing to the design study of mission operation and laser system specifications.
Numerical simulations are performed for two test cases to verify the control law under uncertainty of thrust magnitude and orbital perturbations.

%
The rest of this paper is organized as follows.  Section 2 introduces the spacecraft dynamics including the orbital motion and relative motion.  In addition, the definition and advantages of ROE are presented.  In Section 3, the relative equations of motion between the chaser and the target in powered flight is derived.  First, the linear dynamics of relative motion and the time variation of the orbital elements are formulated.  Next, the analytical solution of the relative motion in powered flight is derived.  In Section 4, this paper proposes the control law for the simultaneous deorbit using the analytical solution.  Section 5 presents the numerical simulation results and verifies the performance of the proposed control law.  Finally, Section 6 concludes this study and describes future works.

\section{Preliminaries}

\subsection{Orbital Motion}

\begin{figure}[tb]
\centering
\includegraphics[scale = 0.43]{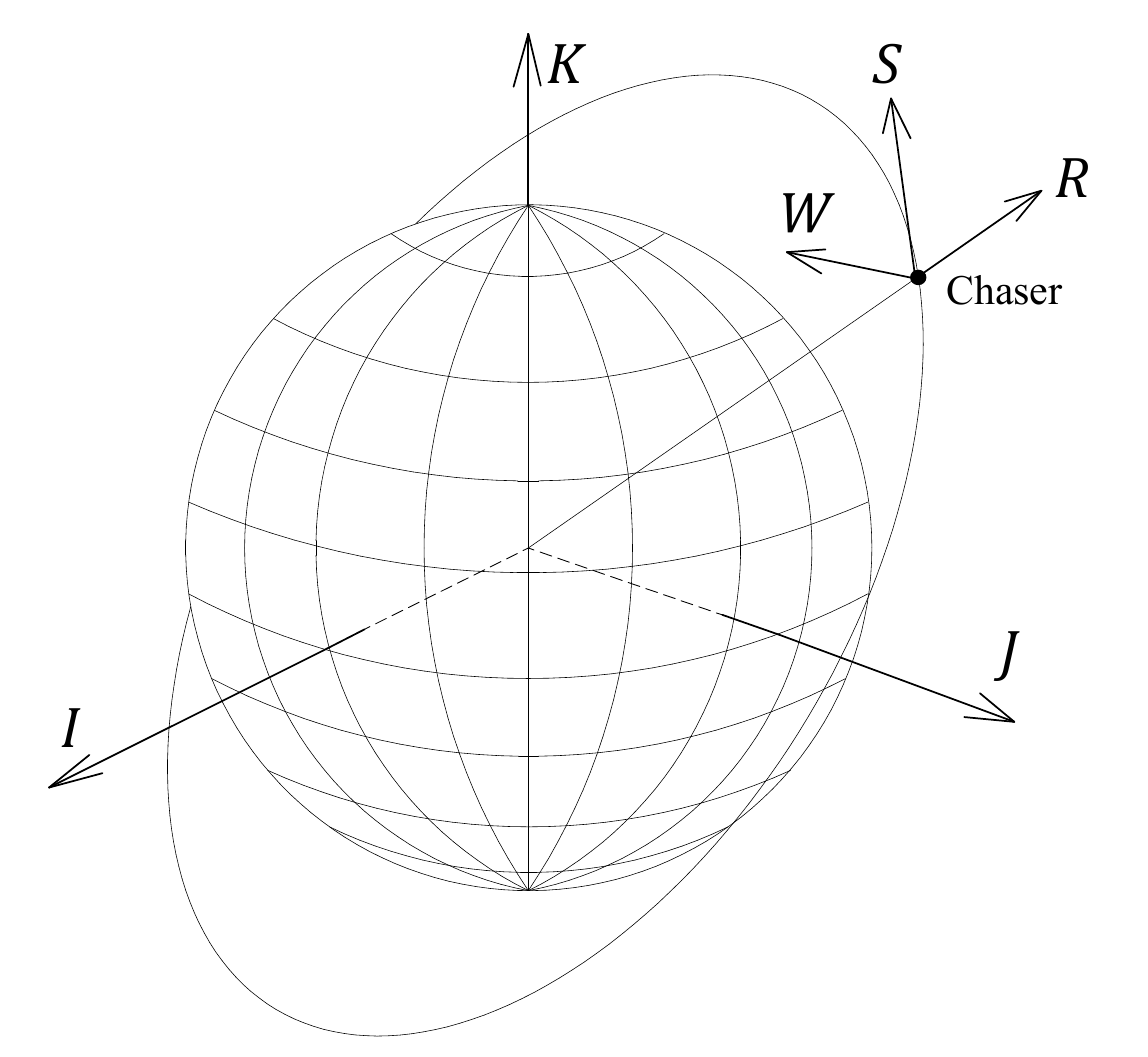}
\caption{Inertial frame and Hill frame}
\label{fig_IJK_RTN}
\end{figure}

%
This paper uses two coordinate frames as illustrated in Fig.~\ref{fig_IJK_RTN}. The inertial frame $\{I,~J,~K\}$ has its origin at the Earth’s center. The $I$ axis is along the vernal equinox direction, the $K$ axis is along the rotational axis of the Earth, and the $J$ axis completes the right-handed frame.  The Hill frame~\citep{alfriend2009spacecraft, vallado2001fundamentals} represents the relative position of a target with respect to a chaser satellite.  Its origin is at the chaser position and its orientation is given by the vector triad $\{R,~T,~N\}$.  The $R$ axis is aligned with the radial direction (positive outward) and the $N$ axis is parallel to the orbital angular momentum vector of the chaser (positive in the orbit normal direction). The $T$ axis completes the right-handed frame.

%
The GVE~\citep{vallado2001fundamentals,larbi2016spacecraft} describe the variation of the Keplerian orbital elements due to accelerations.  
Under the assumption of a near-circular orbit, the GVE are written as follows.
\begin{align}
  \frac{da}{dt} &= \frac{2}{n} \frac{F_T}{m} \label{eq_gauss_a}\\
  \frac{du}{dt} &= n -\frac{2}{n a} \frac{F_R}{m}  -\frac{\sin{u}}{n a \tan{i}} \frac{F_N}{m} \label{eq_gauss_u}\\
  \frac{d e_x}{dt} &= \frac{2 \cos{u}}{n a} \frac{F_T}{m} +\frac{\sin{u}}{n a} \frac{F_R}{m} \label{eq_gauss_ex}\\
  \frac{d e_y}{dt} &= \frac{2 \sin{u}}{n a} \frac{F_T}{m} -\frac{\cos{u}}{n a} \frac{F_R}{m} \label{eq_gauss_ey}\\
  \frac{di}{dt} &= \frac{\cos{u}}{n a} \frac{F_N}{m} \label{eq_gauss_i}\\
  \frac{d \Omega}{dt} &= \frac{\sin{u}}{n a \sin{i}} \frac{F_N}{m} \label{eq_gauss_Omega}
\end{align}where $a$, $i$, $\Omega$, and $u$ indicate the semi-major axis, inclination, right ascension of the ascending node, and the mean argument of latitude, respectively.  Variables $e_x$ and $e_y$ represent the components of the eccentricity vector and are defined as $e_x = e \cos \omega$ and $e_y = e \sin \omega$.  Also, $F_{R}$, $F_{T}$, $F_{N}$ and $m$ are the external forces in the Hill frame and the mass of the satellite, respectively.  In this study, the linear relative equations of motion in powered flight are derived from the GVE.  

\subsection{Relative Motion}

%
Consider two satellites orbiting the Earth.  Let $\bm{r}_c$ and $\bm{\rho}$ be the position of a chaser satellite and the relative position of a target satellite with respect to the chaser satelite in the inertial frame.  The orbital motion of the chaser and target satellites is described as
\begin{equation}
\label{eq_om_c}
\ddot{\bm{r}}_c = -\frac{\mu}{\|\bm{r}_c\|^3}\bm{r}_c + \frac{\bm{F}_{c}}{m_c}
\end{equation}\begin{equation}
\label{eq_om_d}
\ddot{\bm{r}}_c + \ddot{\bm{\rho}} = -\frac{\mu}{\|\bm{r}_c+\bm{\rho}\|^3}(\bm{r}_c+\bm{\rho}) + \frac{\bm{F}_{t}}{m_t}
\end{equation}where $\mu$ is the gravitational constant of the Earth, $\bm{F}_{c}$ and $\bm{F}_{t}$ are the external force of the chaser and target satellites, respectively, and $m_{c}$ and $m_{t}$ are the mass of the chaser and target satellites, respectively.  Taking the difference of these equations yields
\begin{equation}
\label{eq_Hill1}
\ddot{\bm{\rho}} + \frac{\mu}{\|\bm{r}_c+\bm{\rho}\|^3}(\bm{r}_c+\bm{\rho}) - \frac{\mu}{\|\bm{r}_c\|^3}\bm{r}_c = \frac{\bm{F}_{t}}{m_t} - \frac{\bm{F}_{c}}{m_c}
\end{equation}Now, let us assume that the chaser satellite is in a circular orbit and that the distance between the chaser and target is much smaller than the orbital radius of the chaser.  The relative position $\bm{\rho}$ can be expressed through the Hill coordinates as $ \bm{\rho}= x \bm{R} + y \bm{T} + z \bm{N} $.  Since the Hill frame is the rotating frame, the equations of relative motion in the Hill frame are given by
\begin{equation}
\label{eq_Hill}
    \begin{split}
    \ddot{x} -2 n \dot{y} -3 n ^2 x &= \frac{F_{Rt}}{m_t} - \frac{F_R}{m_c} \\
    \ddot{y} +2 n ^2 \dot{x} &= \frac{F_{Tt}}{m_t} - \frac{F_T}{m_c} \\
    \ddot{z} + n ^2 z &= \frac{F_{Nt}}{m_t} - \frac{F_N}{m_c} 
    \end{split}
\end{equation}where $ n $ is the mean motion and the external force $\bm{F}_{c}$ and $\bm{F}_{t}$ are expressed in the Hill frame as $\bm{F}_{c} = [F_{R}, F_{T}, F_{N}]^T$ and $\bm{F}_{t} = [F_{Rt}, F_{Tt}, F_{Nt}]^T$, respectively.  Equation~\eqref{eq_Hill} is called the Hill--Clohessy--Wiltshire equations of relative motion~\citep{clohessy1960terminal}.

\subsection{Relative Orbital Elements}

\begin{figure*}[tb]
\centering
\includegraphics[scale = 0.45]{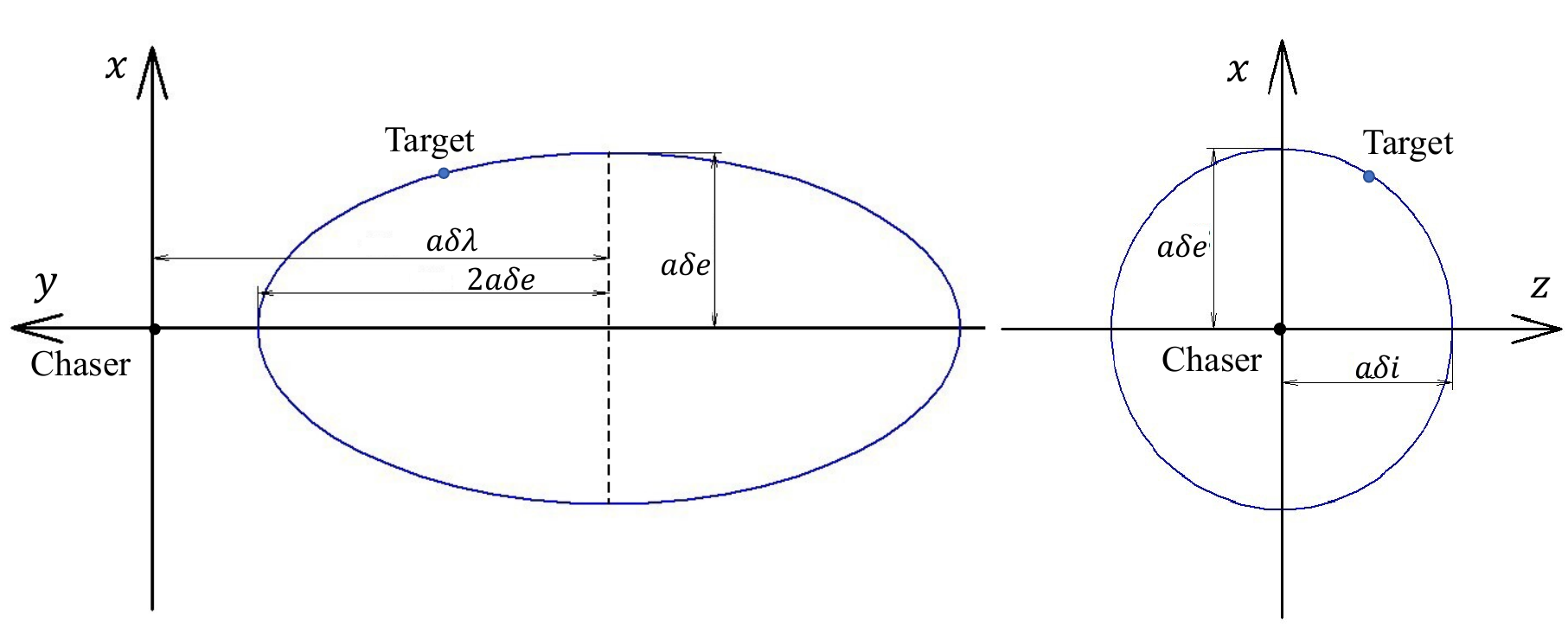}
\caption{Relative motion using ROE}
\label{fig_RM_ROEs}
\end{figure*}

%
The relative motion of a target satellite with respect to a chaser satellite can be parameterized using ROE~\citep{d2010autonomous} as
\begin{equation}
\label{eq_roe}
\delta \bm{\alpha} = \begin{bmatrix} \delta a \\ \delta \lambda \\ \delta e_{x} \\ \delta e_{y} \\ \delta i_{x} \\ \delta i_{y} \end{bmatrix} = \begin{bmatrix} (a_{t} - a)/a \\ (u_{t} - u)+(\Omega_{t} - \Omega)\cos i \\ e_{xt} - e_{x} \\ e_{yt}-e_{y} \\ i_{t}-i \\ (\Omega_{t} - \Omega)\sin i \end{bmatrix}
\end{equation}where the variables with the subscript ``$t$'' refer to those of the target, and the variables without a subscript refer to those of the chaser.  
In ROE, $\delta a $ indicates that the difference in the semi-major axis, and $\delta a = 0$ means a bounded relative motion. The relative mean longitude is expressed with $\delta\lambda$. The relative eccentricity $\delta\bm{e}$ and inclination vectors $\delta\bm{i}$ are defined as follows in the Cartesian and polar representations, respectively
\begin{equation}
\label{eq_e_v}
\delta \bm{e} = 
\begin{bmatrix} \delta e_{x} \\ \delta e_{y}\end{bmatrix} = 
\begin{bmatrix} e_{xt} - e_{x} \\ e_{yt}-e_{y} \end{bmatrix} = 
\delta e \begin{bmatrix} \cos \varphi \\ \sin \varphi \end{bmatrix}
\end{equation}\begin{equation}
\label{eq_i_v}
\delta \bm{i} = 
\begin{bmatrix} \delta i_{x} \\ \delta i_{y}\end{bmatrix} = 
\begin{bmatrix} i_{t}-i \\ (\Omega_{t} - \Omega)\sin i \end{bmatrix} = 
\delta i \begin{bmatrix} \cos \vartheta \\ \sin \vartheta \end{bmatrix}
\end{equation}where $\delta e= \|\delta\bm{e}\|$, $\delta i=\|\delta\bm{i}\|$ , and $\varphi$ and $\vartheta$ are the phases of the relative eccentricity and inclination vectors.  Also, $\delta a$, $\delta\lambda$, and $\delta\bm{e}$ describe the in-plane motion, and $\delta\bm{i}$ describes the out-of-plane motion. 
Note that target orbital elements can be reformulated in terms of chaser orbital elements and ROE using the following identities~\citep{sullivan2016improved}
\begin{align}
\label{eq_oe2oe_d_a} a_{t} &= a(\delta a + 1) \\ 
\label{eq_oe2oe_d_i} i_{t} &= i + \delta i_x \\ 
\label{eq_oe2oe_d_O} \Omega_{t} &= \Omega + \frac{\delta i_y}{\tan i}  \\ 
\label{eq_oe2oe_d_e} e_{t} &= \sqrt{(e_x + \delta e_x)^2+(e_y + \delta e_y)^2} \\ 
\label{eq_oe2oe_d_o} \omega_{t} &= \arctan \left(\frac{e_y + \delta e_y}{e_x + \delta e_x}\right) \\ 
\label{eq_oe2oe_d_u1} u_{t} &= u + \delta \lambda - (\Omega_t - \Omega) \cos i \\ 
\label{eq_oe2oe_d_u2} &= u + \delta \lambda - \frac{\delta i_y}{\tan i} 
\end{align}
%
ROE has four advantages~\citep{DAmico2005}: first, the geometry of the relative motion is immediately visible in terms of the differences in orbital elements.  Second, collision avoidance between satellites can be handled by $e/i$-vector separation~\citep{eckstein1989colocation}.  Third, relative orbit control can be planned using the GVE.  Fourth, the oblateness effects of the Earth and higher-order geopotential effects can be incorporated into the analytical solution of ROE through the convenient description of orbital elements.

\subsection{Transformation matrix of ROE--Hill frame}

%
The relative position and velocity of the target with respect to the chaser can be expressed through the Hill coordinates as $ \bm{x}=[x, y, z, v_x, v_y, v_z]^T $.  The conversion between ROE and relative position and velocity is obtained using the following transformation matrix~\citep{d2010autonomous, gaias2021trajectory}.
\begin{equation}
\label{eq_roes2rm}
\bm{x}(t) = T(t) \delta \bm{\alpha} (t) \\
\end{equation}\begin{equation}
\label{eq_roes2rm_T}
T(t) = 
a \begin{bmatrix} 1 & 0 & -c_{u} & -s_{u} & 0 & 0\\  0 & 1 & 2s_{u} & -2c_{u} & 0 & 0\\  0 & 0 & 0 & 0 & s_{u} & -c_{u} \\ 0 & 0 & n s_{u} & - n c_{u} & 0 & 0\\  -1.5 n & 0 & 2 n c_{u} & 2 n s_{u} & 0 & 0\\  0 & 0 & 0 & 0 & n c_{u} & n s_{u} \end{bmatrix}
\end{equation}where $s_{(\cdot)}$ and $c_{(\cdot)}$ denote $\sin (\cdot)$ and $\cos (\cdot)$, respectively. 

Making the use of the polar representation of the relative eccentricity and inclination vectors, Eqs.~\eqref{eq_roes2rm} and~\eqref{eq_roes2rm_T} become
\begin{equation}
\label{eq_roe2rm_3}
\begin{matrix} 
x &= &a\delta a & &-a\delta e\cos (u-\varphi)  \\
y &= & & a\delta \lambda &-2a\delta e\sin (u-\varphi)  \\
z &= & & &a\delta i\sin (u-\vartheta) \\
v_{x}  &= & & &na\delta e\sin (u-\varphi)  \\
v_{y}  &= &-1.5na\delta a & &+2na\delta e\cos (u-\varphi)  \\
v_{z}  &= & & & na\delta i\cos (u-\vartheta) \end{matrix}
\end{equation}From Eq.~\eqref{eq_roe2rm_3}, the bounded relative orbit ($\delta a = 0$) of the target with respect to the chaser is an ellipse of the semi-major axis $2a\delta e$ in the along-track direction and the semi-minor axis $a\delta e$ in the radial direction as illustrated in Fig.~\ref{fig_RM_ROEs}.  Furthermore, $a \delta \lambda$ and $\varphi$ define the center of the ellipse and the relative pericenter, respectively.  The relative motion of the cross-track is described by a harmonic oscillation of amplitude $a\delta i$  and phase angle $u-\vartheta$.

\section{Equations of relative motion in powered flight}

As the first step to derive the control law for the simultaneous deorbit by laser, this paper derives the analytical solution of relative equations of motion in powered flight using ROE.

\subsection{Formulation}

The linear dynamics of relative motion in powered flight is described using the GVE.  The chaser orbit is assumed to be a near-circular orbit, and the first-order Taylor expansion yields the linear dynamics.

%
The time derivative of Eq.~\eqref{eq_roe} yields~\citep{guffanti2017long}
\begin{equation}
\label{eq_dif_roe}
\delta \dot{\bm{\alpha}} (t) 
=\bm{\zeta}(\delta\bm{\alpha},\bm{a}) = \begin{bmatrix} (\dot{a}_{t} - \dot{a})/a - \dot{a} \delta a / a \\ (\dot{u}_{t} - \dot{u}) +(\dot{\Omega}_{t} - \dot{\Omega}) c_{i} - \dot{i}\delta i_y \\ \dot{e}_{xt} - \dot{e}_{x} \\ \dot{e}_{yt}-\dot{e}_{y} \\ \dot{i}_{t}-\dot{i} \\ (\dot{\Omega}_{t} - \dot{\Omega}) s_{i} + \dot{i} \delta i_y / \tan i \end{bmatrix}
\end{equation}where the acceleration vector $\bm{a}$ is expressed in the Hill frame as $\bm{a} = [F_{Rt}/m_t, F_{Tt}/m_t, F_{Nt}/m_t, F_{R}/m_c, F_{T}/m_c, F_{N}/m_c]^T$.  Substituting the GVE in Eqs.~\eqref{eq_gauss_a}--\eqref{eq_gauss_Omega} into Eq.~\eqref{eq_dif_roe} and performing a first-order Taylor expansion of $\bm{\zeta}$ around the chaser orbit ($\delta\bm{\alpha}=0$) yield the following linear dynamics (See~\ref{Appendix_nonlinear} for a detailed formulation without linearization). 
\begin{align}
\label{eq_form_dif} 
\delta \dot{\bm{\alpha}} (t) &= \left.\frac{\partial\bm{\zeta}}{\partial \bm{\alpha}}\right|_{\delta\bm{\alpha}=0}   \delta \bm{\alpha} + \left.\frac{\partial\bm{\zeta}}{\partial \bm{a}}\right|_{\delta\bm{\alpha}=0} \bm{a} \\
&= A \delta \bm{\alpha} + B \bm{a} \\
\label{eq_form_difA} 
A &= \begin{bmatrix} 0 & 0 & 0 & 0 & 0 & 0\\  -\frac{3}{2} n & 0 & 0 & 0 & 0 & 0\\  0 & 0 & 0 & 0 & 0 & 0\\ 0 & 0 & 0 & 0 & 0 & 0\\  0 & 0 & 0 & 0 & 0 & 0\\  0 & 0 & 0 & 0 & 0 & 0 \end{bmatrix}\\
\label{eq_form_difB} 
B &= \frac{1}{a n}\begin{bmatrix} 0 & 2 & 0 & 0 & -2 & 0\\ -2 & 0 & 0 & 2 & 0 & 0\\ s_{u} & 2 c_{u} & 0 & -s_{u} & -2 c_{u}& 0\\ - c_{u} & 2 s_{u} & 0 & c_{u} & -2 s_{u} & 0\\ 0 & 0 & c_{u} & 0 & 0 & -c_{u} \\ 0 & 0 & s_{u} & 0 & 0 & -s_{u} \end{bmatrix} 
\end{align}Note that the orbital elements in Eqs.~\eqref{eq_form_dif}--\eqref{eq_form_difB} vary with the maneuvers of the satellites, and their changes need to be taken into account for the analytical solution.

\subsection{Analytical solution}

In this section, the analytical solution of relative motion in powered flight is described.  First, the time variation of the orbital elements is formulated.  Second, substituting the analytical solution of the orbital elements into the relative motion dynamics, this study derives the analytical solution of the relative motion in powered flight in Eqs.~\eqref{eq_form_dif}--\eqref{eq_form_difB}.

%
Equation~\eqref{eq_gauss_a} is rewritten as
\begin{align}
\frac{da}{dt} &= 2\sqrt{\frac{a^3}{\mu}} \frac{F_T}{m_c}\\
&= \frac{2}{\sqrt{\mu}} a^{\frac{3}{2}} \frac{F_T}{m_c} \label{eq:dadt}
\end{align}Assuming that the magnitude of the satellite maneuver is constant, Eq.~\eqref{eq:dadt} can be integrated as
\begin{align}
&\int_{a_0}^{a} a^{-\frac{3}{2}}\frac{da}{dt} dt = \int_{t_0}^{t} \frac{2}{\sqrt{\mu}} \frac{F_T}{m_c} dt\\
&\Rightarrow a(t) = a_0 \left( 1- \frac{F_T}{a_0 n_0 m_c} (t-t_0) \right) ^{-2} \label{eq_a_form0}
\end{align}where the subscript ``$0$''  represents the values at $t_0$.  The mean motion $n$ can be expressed using Eq.~\eqref{eq_a_form0} as 
\begin{align}
n(t) &= \sqrt{\frac{\mu}{a^3}} = \sqrt{\mu} a^{-\frac{3}{2}}\\
&= \sqrt{\mu}  a_0^{-\frac{3}{2}} \left( 1- \frac{F_T}{a_0 n_0 m_c} (t-t_0) \right) ^{3}\\
&= n_0 \left( 1- \frac{F_T}{a_0 n_0 m_c} (t-t_0) \right) ^{3}\label{eq_n_form0}
\end{align}Equations \eqref{eq_a_form0} and \eqref{eq_n_form0} are expanded about the initial time $t_{0}$ by the first-order Taylor expansion as
\begin{align}
\label{eq_a_form1}
a(t) 
&= a_0 \left( 1- \frac{F_T}{a_0 n_0 m_c} (t-t_0) \right) ^{-2} \\
&\approx a_0 \left( 1 + 2 \frac{F_T}{a_0 n_0 m_c} (t-t_0) \right) \\
&= a_0\left(1 +2J(t-t_0)\right)
\end{align}\begin{align}
\label{eq_n_form1}
n(t) 
&= n_0 \left( 1- \frac{F_T}{a_0 n_0 m_c} (t-t_0) \right) ^{3} \\
&\approx n_0 \left( 1 -3 \frac{F_T}{a_0 n_0 m_c} (t-t_0) \right) \\
&= n_0\left(1 -3J(t-t_0)\right)
\end{align}where
\begin{equation}
\label{eq_J}
J = \frac{1}{a_0 n_0} \frac{F_{T}}{m_c}
\end{equation}
For the further derivation, the following equation is derived by the first-order Taylor expansion of Eqs.~\eqref{eq_a_form1} and~\eqref{eq_n_form1} as
\begin{equation}
\label{eq_an_form1}
\frac{1}{a(t)n(t)} = \frac{1}{a_0 n_0} \left(1 +J(t-t_0)\right)
\end{equation}
%
Assuming that the maneuver is in the in-plane direction, the analytical solution for the mean argument of latitude $u$ can be expressed using Eqs.~\eqref{eq_gauss_u},~\eqref{eq_n_form1} and,~\eqref{eq_an_form1} as 
\begin{align}
\frac{du}{dt} &= n -\frac{2}{n a} \frac{F_R}{m_c} \\
&= n_0\left(1 -3J(t-t_0)\right) - \frac{2}{a_0 n_0} \left(1 +J(t-t_0)\right) \frac{F_R}{m_c} \\
&= \left(n_0 - \frac{2 F_R}{a_0 n_0 m_c}\right) - \left(3J n_0 +\frac{2}{a_0 n_0 m_c}J F_R\right) (t-t_0)
\end{align}Integrating both sides yields
\begin{align}
u(t) - u_0 =& \left(n_0 - \frac{2 F_R}{a_0 n_0 m_c}\right)(t-t_0) \nonumber \\
& - \frac{1}{2} \left(3J n_0 +\frac{2}{a_0 n_0 m_c}J F_R\right) (t-t_0)^2 \\
\Rightarrow u(t) =& W_1(t-t_0) -W_2(t-t_0)^2 +u_0 \label{eq_u_form1}
\end{align}where
\begin{align}
   \label{eq_W1} W_1 &= n_0 -\frac{2}{a_0 n_0} \frac{F_{R}}{m_c} \\ W_2 &= \frac{1}{a_0^2 n_0^2} \frac{F_{R}}{m_c}\frac{F_{T}}{m_c} +\frac{3}{2 a_0} \frac{F_{T}}{m_c}
\end{align}%
Next, the analytical solution of Eq.~\eqref{eq_form_dif} is obtained using the variation of parameters method and Eqs.~\eqref{eq_a_form1},~\eqref{eq_n_form1} and~\eqref{eq_u_form1}.  

First, the homogeneous solution of $\delta a$ can be found from Eq.~\eqref{eq_form_dif}
\begin{align}
\delta \dot{a} (t) &= 0 \\ \delta a (t) &= \delta a_0 \label{eq_Da_ana}
\end{align}Using Eqs.~\eqref{eq_n_form1} and~\eqref{eq_Da_ana}, $ \delta \dot{\lambda}$ in Eq.~\eqref{eq_form_dif} can be analytically integrated as
\begin{align}
\delta \dot{\lambda} (t) &= -\frac{3}{2} n \delta a (t) \\
&= -\frac{3}{2} n_0\left(1 -3J(t-t_0)\right) \delta a_0\\
&\Rightarrow \delta \lambda (t) = -\frac{3}{2} n_0\left( (t-t_0) -\frac{3 J}{2}(t-t_0)^2 \right) \delta a_0 + \delta \lambda_0 \label{eq_Dl_ana}
\end{align}The homogeneous solution of Eq.~\eqref{eq_form_dif} is written as
\begin{equation}
\label{eq_form_homo}
\delta \bm{\alpha} (t) = \Phi(t, t_0, \bm{F}_{c,{\rm in}}/m_c) \delta \bm{ \alpha} (t_0) 
\end{equation}\begin{equation}
\label{eq_formPhi}
\Phi(t, t_0, \bm{F}_{c,{\rm in}}/m_c) =
\begin{bmatrix} 1 & 0 & 0 & 0 & 0 & 0\\ \Phi_{21} & 1 & 0 & 0 & 0 & 0\\  0 & 0 & 1 & 0 & 0 & 0\\ 0 & 0 & 0 & 1 & 0 & 0\\  0 & 0 & 0 & 0 & 1 & 0\\  0 & 0 & 0 & 0 & 0 & 1 \\ \end{bmatrix}
\end{equation}\begin{align}
\Phi_{21} &= -\frac{3}{2}n_0 \left( (t-t_0)-\frac{3}{2}J(t-t_0)^2 \right)
\end{align}\begin{align}
\bm{F}_{c,{\rm in}} &= [F_R, F_T]^T 
\end{align}where $\Phi(\cdot,\cdot,\cdot)$ is the state transition matrix, $\bm{F}_{c,{\rm in}}$ is the in-plane external force of the chaser.  
Note that $J$ is the function of $F_T$ as shown in Eq.~\eqref{eq_J}.  
Since $\bm{F}_{c,{\rm in}}$ is the variable parameter, the arguments of the state transition matrix $\Phi$ include $F_{c,{\rm in}}$ as shown in Eq.~\eqref{eq_formPhi}.

%
An analytical solution of Eq.~\eqref{eq_form_dif} is written as
\begin{align}
\label{eq_form0}
\delta \bm{\alpha} (t) =& \Phi(t, t_0, \bm{F}_{c,{\rm in}}/m_c) \delta \bm{ \alpha} (t_0) \nonumber\\
&+ \int_{t_0}^{t} \Phi(\tau, t_0, \bm{F}_{c,{\rm in}}/m_c) B (\tau) \bm{a} d \tau
\end{align}
The analytical solution for the relative equations of motion between the chaser and the target satellites in powered flight can be expressed as follows (See~\ref{Appendix_transformation} for the detailed transformation).
\begin{align}
\label{eq_form}
\delta \bm{\alpha} (t) =& \Phi(t, t_0, \bm{F}_{c,{\rm in}}/m_c) \delta \bm{ \alpha} (t_0) \nonumber\\
&+ \Psi_c (t, t_0, \bm{F}_{c,{\rm in}}/m_c) \bm{F}_{c,{\rm in}}/m_c + \Psi_t (t, t_0, \bm{F}_{c,{\rm in}}/m_c) \bm{F}_t /m_t
\end{align}where
\begin{align}
\label{eq_formPsiC}
&\Psi_c (t, t_0, \bm{F}_{c,{\rm in}}/m_c) = \nonumber\\
&\frac{1}{a_0 n_0 W_1}\begin{bmatrix} 0 & -\Psi_{21}\\ \Psi_{21} & -\Psi_{22} \\  c_{u} - c_{u_0} & -2(s_{u} - s_{u_0} )\\s_{u} - s_{u_0} & 2(c_{u} - c_{u_0} )\\  0 & 0\\  0 & 0 \end{bmatrix}
\end{align}\begin{align}
\label{eq_formPsiD}
&\Psi_t (t, t_0, \bm{F}_{c,{\rm in}}/m_c) = \nonumber\\
&\frac{1}{a_0 n_0 W_1}\begin{bmatrix} 0 & \Psi_{21} & 0 \\  -\Psi_{21} & \Psi_{22}  & 0 \\  c_{u_0} - c_{u} & 2(s_{u} - s_{u_0} ) & 0\\ s_{u_0} - s_{u} & -2(c_{u} - c_{u_0} ) & 0 \\  0 & 0 & s_{u_0} - s_{u} \\  0 & 0 & c_{u_0} - c_{u} \end{bmatrix}
\end{align}\begin{align}
\Psi_{21} =& W_1[J(t-t_0)+2](t-t_0) \\
\Psi_{22} =& \frac{n_0 W_1}{8} [9J^2(t-t_0)^2+4J(t-t_0)-12](t-t_0)^2
\end{align}
Note that $W_1$ is the function of $\bm{F}_{c,{\rm in}}/m_c$ as shown in Eq.~\eqref{eq_W1}.

\section{Formation keeping strategy}

\begin{figure}[tb]
\centering
\includegraphics[scale = 0.4]{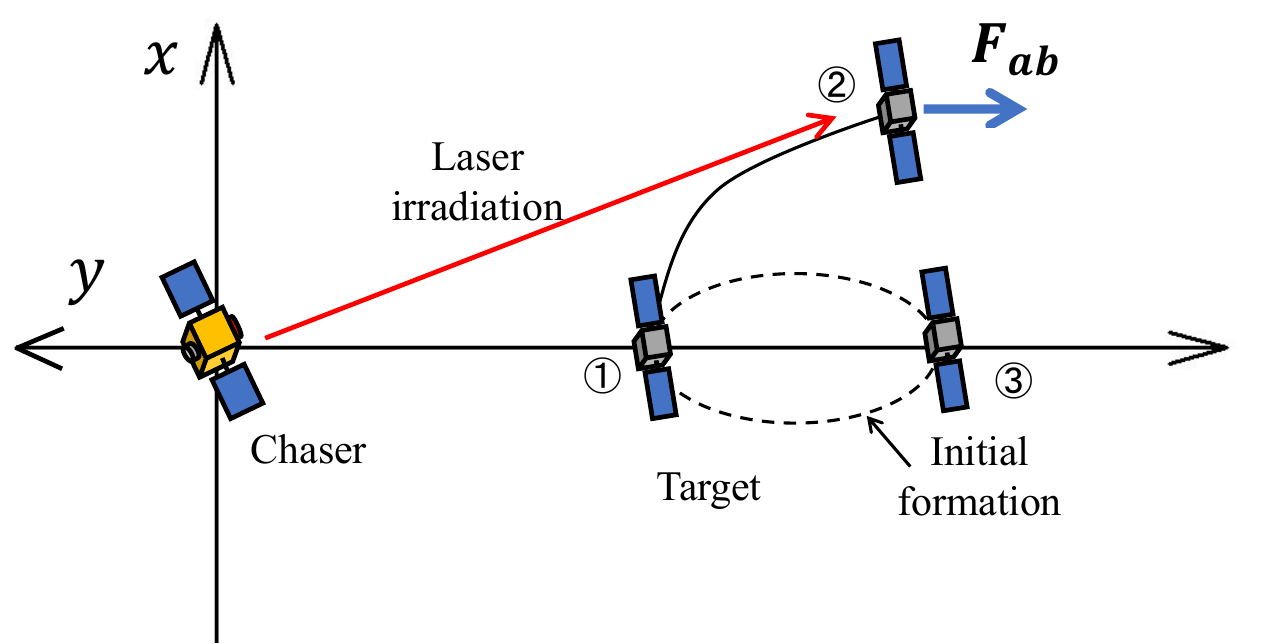}
\caption{The operation of simultaneous deorbit}
\label{fig_InitialFormation2_opp}
\end{figure}

%
Using the analytical solution in Eq.~\eqref{eq_form}, this paper proposes a control law for simultaneous deorbit, which determines the timings and directions of the laser ablation and the electrical thrust so that the formation periodically returns to a desired formation as shown in Fig.~\ref{fig_InitialFormation2_opp}.  By repeating the control law, both the chaser satellite and the target can deorbit while maintaining the relative motion to keep irradiating laser pulses.  
Note that the analytical solution in Eq.~\eqref{eq_form} is derived from linear equations of motion without orbital perturbations. The use of accurate relative motion model~\citep{franzini2020relative} would improve the control performance, which will be included in future work.

\subsection{Assumptions}

\begin{figure}[tb]
\centering
\includegraphics[scale = 0.4]{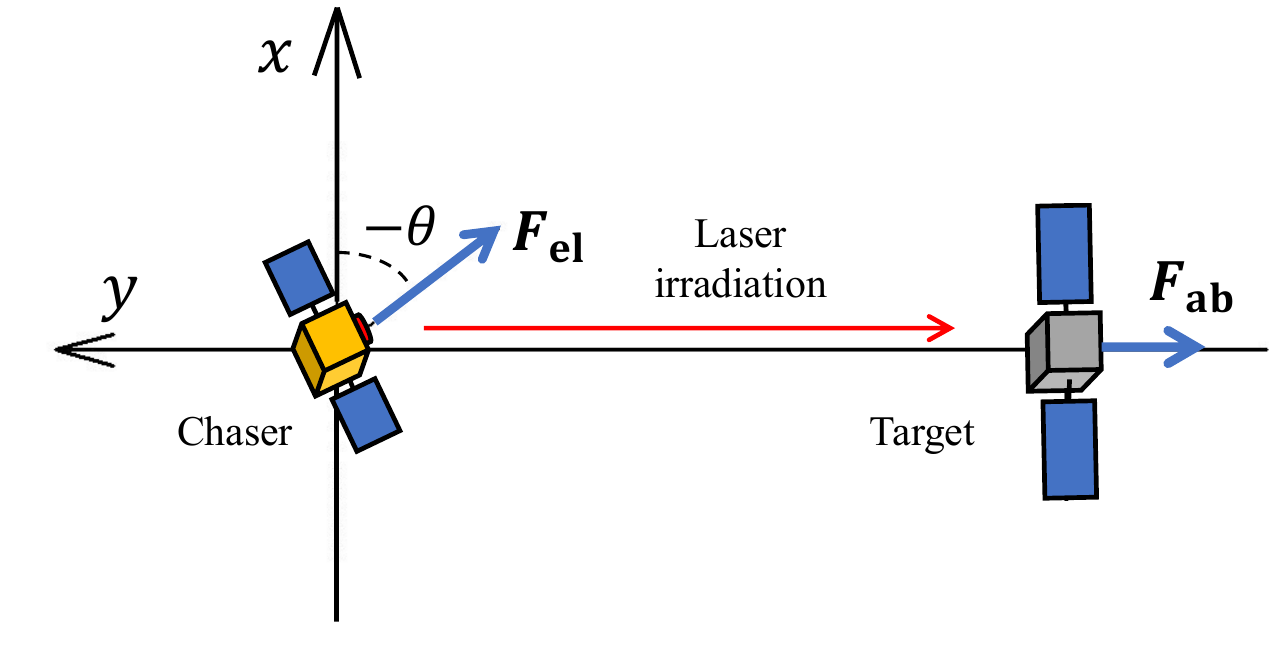}
\caption{Formation keeping for simultaneous deorbit}
\label{fig_ADRmission2_opp}
\end{figure}

%
From the point of view of practical mission constraints and requirements, the following assumptions are made to design the control method for formation keeping.
First, the magnitudes of the ablation force and the electrical thrust are assumed to be constant~\citep{vasile2014improved}.  As shown in Fig.~\ref{fig_ADRmission2_opp}, the ablation force $\bm{F}_{\rm ab}$ is generated along the normal vector of the irradiated surface~\citep{chang2014experimental}.  In this study, the ablation direction is assumed to be the $-T$ direction in the Hill frame, i.e., $\bm{F}_{\rm ab} = [0, -F_{\rm ab}, 0]^T$.  The electrical propulsion thrust $\bm{F}_{\rm el}$ is assumed to be tilted by $\theta$ with respect to the radial direction in the Hill frame, i.e., $\bm{F}_{\rm el} = [F_{\rm el} \cos{\theta}, F_{\rm el} \sin{\theta}]^T$.  In other words, the external forces of the chaser and target are $\bm{F}_{c,{\rm in}} = \bm{F}_{\rm el}$ and $\bm{F}_t = \bm{F}_{\rm ab}$, respectively.
Second, the electrical propulsion and the laser ablation system cannot be operated simultaneously due to power constraints.

\subsection{Control strategy}

\begin{figure}[tb]
\centering
\includegraphics[scale = 0.38]{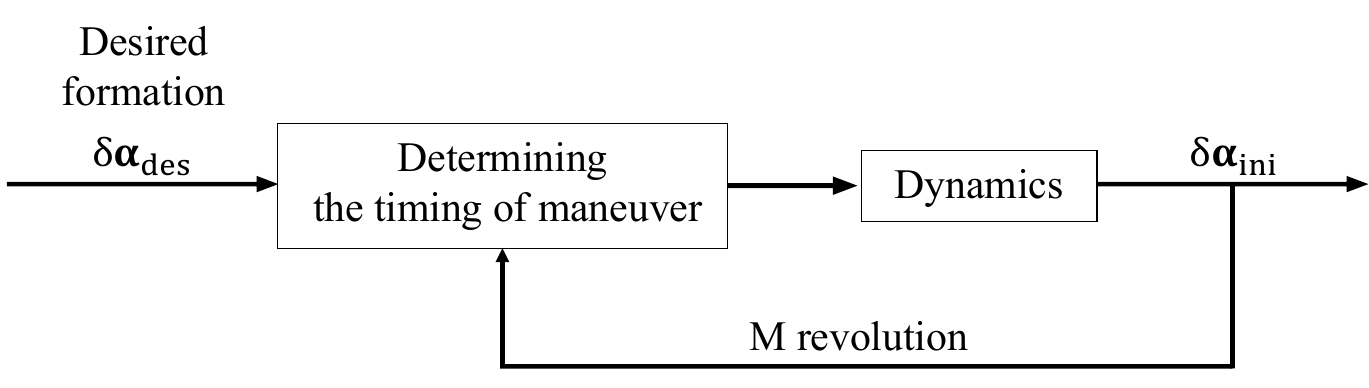} 
\caption{Control flow to compensate the disturbed relative orbit}
\label{fig_ADRmission_control}
\end{figure}

%
Using the analytical solution in Eq.~\eqref{eq_form}, this paper derives the formation keeping control law for simultaneous deorbit.  Although a similar approach can be found in~\cite{di2018continuous}, the proposed control law in the current paper considers both accelerations of chaser and target.  The proposed control law determines the timings and directions of the laser ablation and the electrical thrust so that the formation periodically returns to a desired formation.  The laser ablation and electrical propulsion are assumed to be operated exclusively.  In other words, laser ablation and electrical thrust are operated without overlap.

%
The difference between the initial ROE $\delta \bm{\alpha}_{\rm ini}$ and the desired ROE $\delta \bm{\alpha}_{\rm des}$ at the end of the maneuvering interval is defined as follows.
\begin{equation}
\label{eq_control_des}
\Delta \bm{\alpha}_{\rm des} = \delta \bm{\alpha}_{\rm des} - \Phi(t_m, t_{\rm ini}, \bm{F}_{\rm el}/m_c) \delta \bm{\alpha}_{\rm ini}
\end{equation}where $t_{\rm ini}$ and $t_m$ indicate the initial time and the end time of the maneuvering interval, respectively.  
The maneuvering interval is the period of time to return to a desired formation and is set to one orbital revolution in this paper.  
The state transition matrix $\Phi(\cdot,\cdot,\cdot)$ is defined in Eq.~\eqref{eq_formPhi}.  When laser ablation and electrical propulsion are performed, respectively, $N_{\rm ab}$ and $N_{\rm el}$ times, the following equation is obtained by Eqs.~\eqref{eq_form}--\eqref{eq_formPsiD}.
\begin{align}
\label{eq_control_keeping}
\Delta \bm{\alpha}_{\rm des} =& \sum_{j=1}^{N_{\rm ab}} \Phi(t_m, t_{j, {\rm ab}, f}, \bm{0}) \Psi_t(t_{j,{\rm ab},f},t_{j,{\rm ab},0}, \bm{0}) \bm{F}_{\rm ab}/m_t \nonumber\\ 
&+ \sum_{j=1}^{N_{\rm el}} \Phi(t_m,t_{j,{\rm el},f}, \bm{F}_{\rm el}/m_c) \Psi_c(t_{j,{\rm el},f},t_{j,{\rm el},0}, \bm{F}_{\rm el}/m_c) \bm{F}_{\rm el}/m_c
\end{align}where $t_{j, {\rm ab}, 0}$ and $t_{j, {\rm ab}, f}$ are the start and end time of the $j$-th laser ablation, respectively, whereas $t_{j, {\rm el}, 0}$ and $t_{j, {\rm el}, f}$ are the start and end time of the $j$-th electrical thrust, respectively.  Equation~\eqref{eq_control_keeping} consists of 6 nonlinear equations.  Because this paper considers in-plane formation keeping, the number of Eq.~\eqref{eq_control_keeping} is reduced to four.  In other words, the four design parameters selected from the timings ($t_{j, {\rm ab}, 0}$,$t_{j, {\rm ab}, f}$,$t_{j, {\rm el}, 0}$ and $t_{j, {\rm el}, f}$) and the directions ($\theta$) can be determined using Eq.~\eqref{eq_control_keeping}.

%
In actual satellite operation, there are uncertainties such as the switching interval, the magnitude of electrical thrust and ablation thrust, and the accelerations of orbital perturbations.  They cause a relative position error from the desired relative motion.  This relative position error can be reduced by the proposed control law because Eq.~\eqref{eq_control_des} is updated to compensate for the error for each orbital revolution.  That is, the proposed controller uses the desired formation $\delta \bm{\alpha}_{\rm des}$ and the initial relative position $\delta \bm{\alpha}_{\rm ini} $ to determine the timings and direction of the maneuver that correct the disturbed relative orbit after $M$ revolutions (see Fig.~\ref{fig_ADRmission_control}).  Furthermore, using gain $K\geq 1.0$, Eq.~\eqref{eq_control_des} can be rearranged to improve the disturbance correction as follows.
\begin{equation}
\label{eq_control_des2}
\Delta \bm{\alpha}_{\rm des} = K \left( \delta \bm{\alpha}_{\rm des} - \Phi(t_m, t_{\rm ini}, \bm{F}_{\rm el}/m_c) \delta \bm{\alpha}_{\rm ini} \right)
\end{equation}where the control gain $K$ is empirically determined by varying $K$ in the range of $1.0$ to $2.0$ to match the uncertainties in this paper.  When the uncertainties is not considered, $K =1.0$ is used.

\subsection{Maneuver strategy}

\begin{figure}[tb]
\centering
\includegraphics[scale = 0.4]{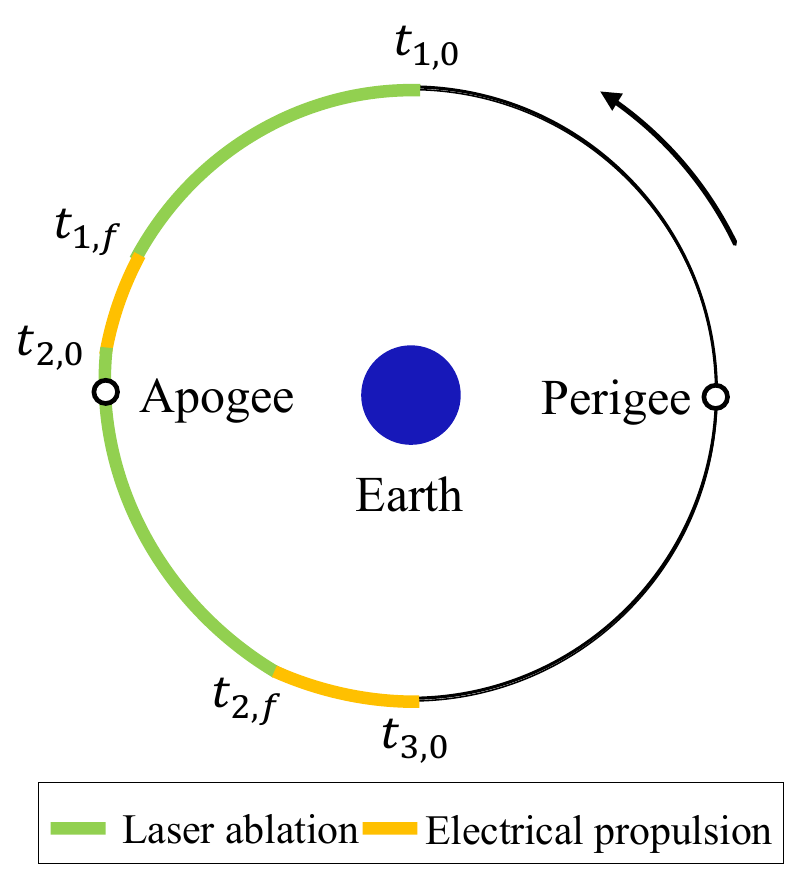}
\caption{Maneuver strategy 1}
\label{fig_MS1}
\end{figure}
\begin{figure}[tb]
\centering
\includegraphics[scale = 0.4]{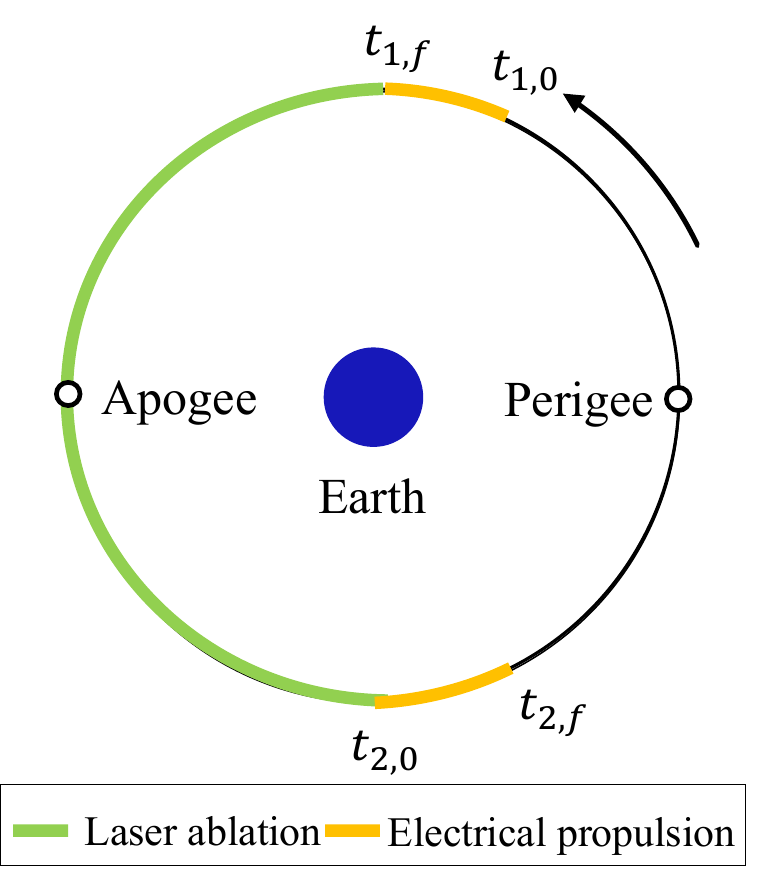}
\caption{Maneuver strategy 2}
\label{fig_MS2}
\end{figure}

%
The proposed control law in Eqs.~\eqref{eq_control_keeping} and~\eqref{eq_control_des2} is flexible in terms of the desired formation and maneuver timings.  Thus, this study considers two maneuver strategies to periodically return to a desired formation in one orbital revolution.

~\citet{fukii2022assessing} revealed that a low-thrust active in 50\% of the orbital period centered about apogee is the most effective for deorbiting the target object in terms of deorbiting time.  Thus, laser ablation and electrical thrust are applied in the vicinity of the apogee in this study.  A geometric illustration of the maneuver strategy 1 is described in Fig.~\ref{fig_MS1}.  The maneuvers are composed of two laser ablation (green) and two electric propulsion (yellow).  The initial time $t_{\rm ini}$ and the end time $t_m$ are assumed to be the time to pass through perigee, respectively.  The first laser ablation start time $t_{1, 0}$ and the second electric propulsion end time $t_{3,0}$ are assumed to be the time to pass through the position of mean argument of latitude $u=1/2\pi$ and $u=3/2\pi$, respectively.  On the other hand, the timings $t_{1, f}$, $t_{2, 0}$, $t_{2, f}$, and the direction of electrical propulsion $\theta$ are the design parameters of the proposed control law in Eq.~\eqref{eq_control_keeping}.  Considering this maneuver strategy 1, Eq.~\eqref{eq_control_keeping} can be rearranged so that
\begin{align}
\label{eq_control_keeping_MS1}
\Delta \bm{\alpha}_{\rm des} =& \Phi(t_m, t_{1,f}, \bm{0}) \Psi_t(t_{1,f},t_{1,0}, \bm{0}) \bm{F}_{\rm ab}/m_t \nonumber\\ 
& + \Phi(t_m,t_{2,0}, \bm{F}_{\rm el}/m_c) \Psi_c(t_{2,0},t_{1,f}, \bm{F}_{\rm el}/m_c) \bm{F}_{\rm el}/m_c \nonumber\\ 
& + \Phi(t_m,t_{2,f}, \bm{0}) \Psi_t(t_{2,f},t_{2,0}, \bm{0}) \bm{F}_{\rm ab}/c_t \nonumber\\ 
& + \Phi(t_m,t_{3,0}, \bm{F}_{\rm el}/m_c) \Psi_c(t_{3,0},t_{2,f}, \bm{F}_{\rm el}/m_c) \bm{F}_{\rm el}/m_c
\end{align}In this maneuver strategy, the design parameters $t_{1, f}$, $t_{2, 0}$, $t_{2, f}$ and $\theta$ are determined by solving Eqs.~\eqref{eq_control_des2} and~\eqref{eq_control_keeping_MS1} using Newton iteration.

In the long-term operation of the laser ADR, the number of switches between the laser system and electric propulsion could be a critical factor.  Maneuver strategy 2 has fewer switches than maneuver strategy 1 as shown in Fig.~\ref{fig_MS2}.  The maneuver strategy 2 consists of one laser ablation (green) and two electrical propulsions (yellow).  The laser ablation start time $t_{1, f}$ and end time $t_{2,0}$ are assumed to be the time to pass through the position of 50\% of the orbital period centered about the apogee, respectively.  On the other hand, the electrical propulsion timings $t_{1, 0}$, $t_{2, f}$, the first electrical propulsion direction $\theta_1$, and the second electrical propulsion direction $\theta_2$ are the design parameters in Eq.~\eqref{eq_control_keeping}.  In the maneuver strategy 2, Eq.~\eqref{eq_control_keeping} can be rearranged so that
\begin{align}
\label{eq_control_keeping_MS2}
\Delta \bm{\alpha}_{\rm des} =& \Phi(t_m,t_{1,f}, \bm{F}_{{\rm el}, 1}/m_c) \Psi_c(t_{1,f},t_{1,0}, \bm{F}_{{\rm el}, 1}/m_c) \bm{F}_{{\rm el}, 1}/m_c\nonumber\\ 
& + \Phi(t_m,t_{2,0} \bm{0}) \Psi_t(t_{2,0},t_{1,f}, \bm{0}) \bm{F}_{\rm ab}/m_t\nonumber\\ 
& + \Phi(t_m,t_{2,f}, \bm{F}_{{\rm el}, 2}/m_c) \Psi_c(t_{2,f},t_{2,0}, \bm{F}_{{\rm el}, 2}/m_c) \bm{F}_{{\rm el}, 2}/m_c
\end{align}where
\begin{align}
\label{eq_control_keeping_MS2_supp1}
\bm{F}_{{\rm el}, 1} &= [F_{\rm el} c_{\theta_1}, F_{\rm el} s_{\theta_1}]^T \\ 
\label{eq_control_keeping_MS2_supp2}
\bm{F}_{{\rm el}, 2} &= [F_{\rm el} c_{\theta_2}, F_{\rm el} s_{\theta_2}]^T
\end{align}Similarly, the design parameters $t_{1, 0}$, $t_{2, f}$, $\theta_1$, and $\theta_2$ are determined by solving Eqs.~\eqref{eq_control_des2} and~\eqref{eq_control_keeping_MS2} using Newton iteration.

The proposed control method allows one to select the various design parameters.  This flexibility of the control law will contribute to the design study of mission operation and laser system specifications.  In the next section, numerical simulations are performed to verify the control law using these two maneuver strategies.

\section{Numerical simulation}

\subsection{Simulation condition}

\begin{table}[tb]
    \caption{Simulation condition}
    \label{table_con}
    \centering
    \begin{tabular}{lc}
    \hline \hline
    Geopotential &  $8\times8$\\ 
    Target mass, $m_{t}$ & 150 [kg] \\
    Chaser mass, $m_{c}$ & 150 [kg] \\
    Electric thrust, $F_{\rm el}$ & 10 [mN] \\
    Laser ablation force, $F_{\rm ab}$ & 0.72 [mN] \\
    Magnitude uncertainty & HND \\
    mean and standard deviation $(\bar{X}, \sigma)$ & (0.0, 0.05) \\
    \hline \hline
    \end{tabular}
\end{table}
\begin{table}[tb]
    \caption{Chaser satellite orbital element}
    \label{table_OE}
    \centering
    \begin{tabular}{cccccc}
    \hline \hline
    $a~\rm{[km]}$ & $e$ & $i ~\rm{[deg]}$ & $\Omega ~\rm{[deg]}$ & $\omega ~\rm{[deg]}$ & $u ~\rm{[deg]}$ \\
    \hline
    7578.14 & 0 & 87.9 & 0 & 0 & 0\\ 
    \hline \hline
    \end{tabular}
\end{table}
\begin{table}[tb]
    \caption{Desired formation}
    \label{table_ROE}
    \centering
    \begin{tabular}{cccccc}
    \hline \hline
    $a \delta a \rm{[m]}$ & $a \delta \lambda \rm{[m]}$ & $a \delta e_x \rm{[m]}$ & $a \delta e_y \rm{[m]}$ & $a \delta i_x \rm{[m]}$ & $a \delta i_y \rm{[m]}$ \\
    \hline
    0 & -100 & 15 & 0 & 15 & 0\\ 
    \hline \hline
    \end{tabular}
\end{table}

%
Numerical simulations are performed for two test cases to verify the control law under the uncertainty of thrust magnitude. 
The simulation conditions are summarized in Table~\ref{table_con}.  The laser ablation force is assumed to be $0.72~\rm{mN}$ without any uncertainty, supposing a $20~\rm{\mu Ns/J}$ impulse and a laser power of $1~\rm{J}$ at $36~\rm{Hz}$~\citep{tsuno2020impulse}.  The initial orbital elements of the chaser are summarized in Table~\ref{table_OE}.  
The initial orbit elements of the target are computed using the identities in Eqs.~\eqref{eq_oe2oe_d_a}--\eqref{eq_oe2oe_d_u2}.  The proposed controller aims to return to the formation in Table~\ref{table_ROE} after one revolution.  In this simulation, the initial ROE is the same as the 
formation for simplicity.  The proposed control law can be applied to any desired formation.  This paper selected a safety ellipse~\citep{shuster2021analytic} in Table~\ref{table_ROE} that can be achieved by a bounded relative motion ($\delta a = 0$) and a parallel or antiparallel alignment of the relative eccentricity and inclination vectors ($\delta\bm{e} \parallel \delta\bm{i}$).  This is passively safe because the target does not intersect the tangential axis in an unperturbed relative motion.  
Note that this desired formation in this paper does not consider the safety requirements and constraints of the system in detail.  In actual mission, it is essential to take into account the collision avoidance requirement and laser system specification such as the laser focal length, the laser irradiating angle, and the camera angle.

%
The Cowell's formulation is propagated to obtain the history of the position and velocity of the chaser and the target~\citep{vallado2001fundamentals,di2018continuous}.  The numerical propagator includes the oblateness effect of the Earth and the uncertainties of ablation thrust.  The magnitude uncertainty of the force is defined as the multiplicative uncertainty.  The magnitude of the ablation force with uncertainty $\| \hat{\bm{F}}_{\rm ab} \|$ is written as
\begin{align}
    \|\hat{\bm{F}}_{\rm ab}\| = \eta \|\bm{F}_{\rm ab}\|
\end{align}where $\eta$ is the magnitude uncertainty coefficient and is assumed to follow the half-normal distribution (HND)~\citep{cooray2008generalization}.  The HND is a probability distribution truncated in the interval given by upper bound.  The probability density function of the HND is given by
\begin{align}
    p(X) = \left\{  
    \begin{matrix} \sqrt{\frac{2}{\pi}}\frac{1}{\sigma} \exp \left(-\frac{1}{2} \left(\frac{X-\overline{X}}{\sigma} \right)^2 \right) &\rm{if} &X \geq 0\\
    0 &\rm{if} &X < 0 \end{matrix} \right.
\end{align}where $\overline{X}$ is the mean and $\sigma$ is the standard deviation.  Using the HND, the magnitude of the ablation force does not become larger than the ideal value of $0.72~\rm{mN}$ in Table~\ref{table_con}.
The control gain $K$ in Eq.~\eqref{eq_control_des2} is set to $K = 1.5$ considering the simulation conditions and the uncertainties of the ablation force in Table~\ref{table_con}.

\subsection{Test case 1}

\begin{table}[tb]
    \caption{Design parameters for test case 1}
    \label{table_TC2}
    \centering
    \begin{tabular}{ccccc}
    \hline \hline
    Revolution, $M$ & $t_{1, f} ~\rm{[s]}$ & $t_{2, 0} ~\rm{[s]}$ & $t_{2, f} ~\rm{[s]}$ & $\theta ~\rm{[deg]}$ \\
    \hline
    1 & 2818.2 & 2974.8 & 4825.3 & 209.7\\ 
    2 & 9048.9 & 9231.6 & 11421 & 145.2\\ 
    3 & 15951 & 16103 & 17957 & 209.8\\ 
    4 & 22286 & 22451 & 24555 & 157.6\\ 
    \hline \hline
    \end{tabular}
\end{table}
\begin{figure}[tb]
\centering
\includegraphics[scale = 0.3]{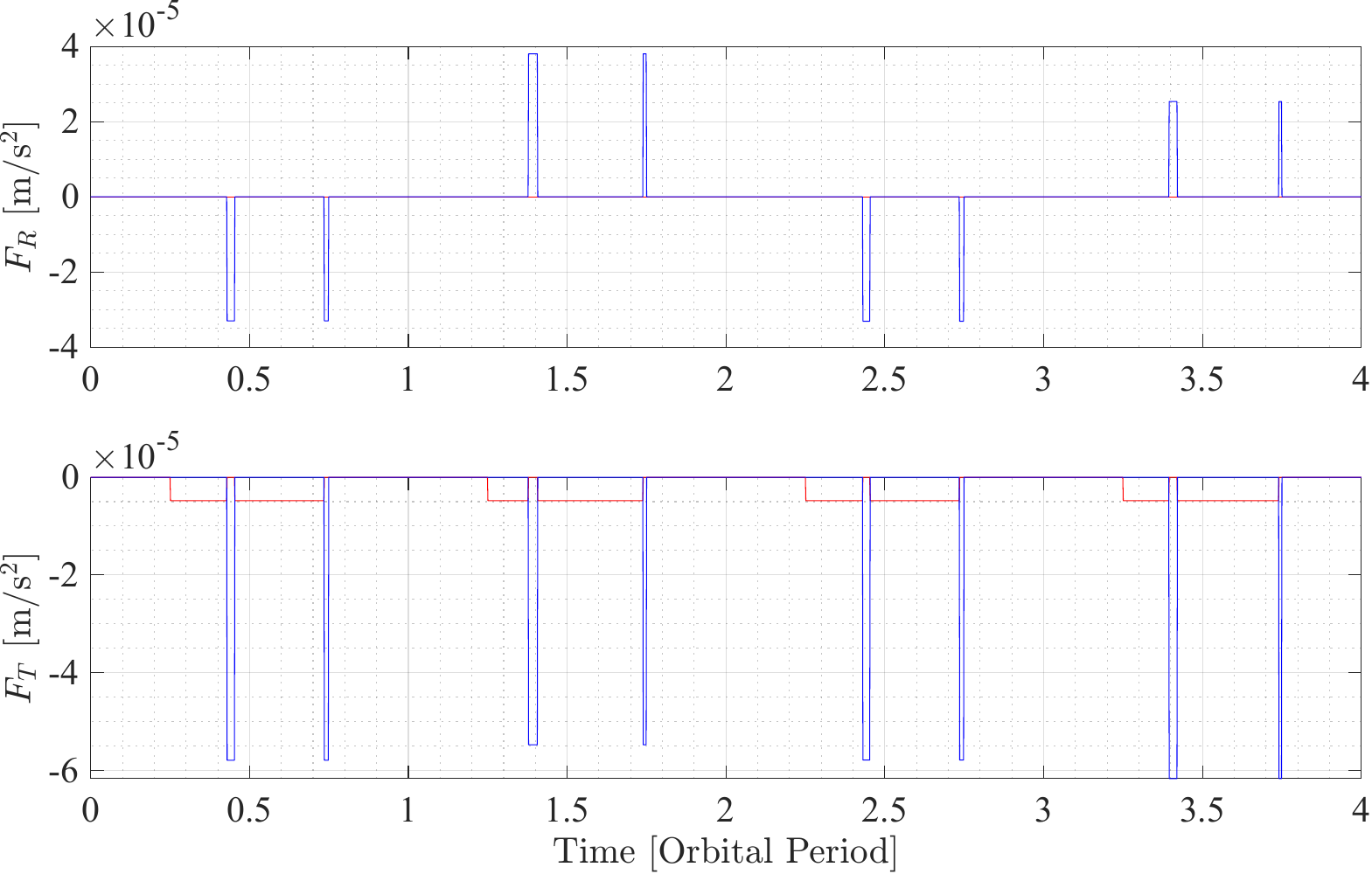}
\caption{Maneuver timing for test case 1}
\label{fig_TC2_R_timing}
\end{figure}
\begin{figure*}[tb]
\centering
\includegraphics[scale = 0.45]{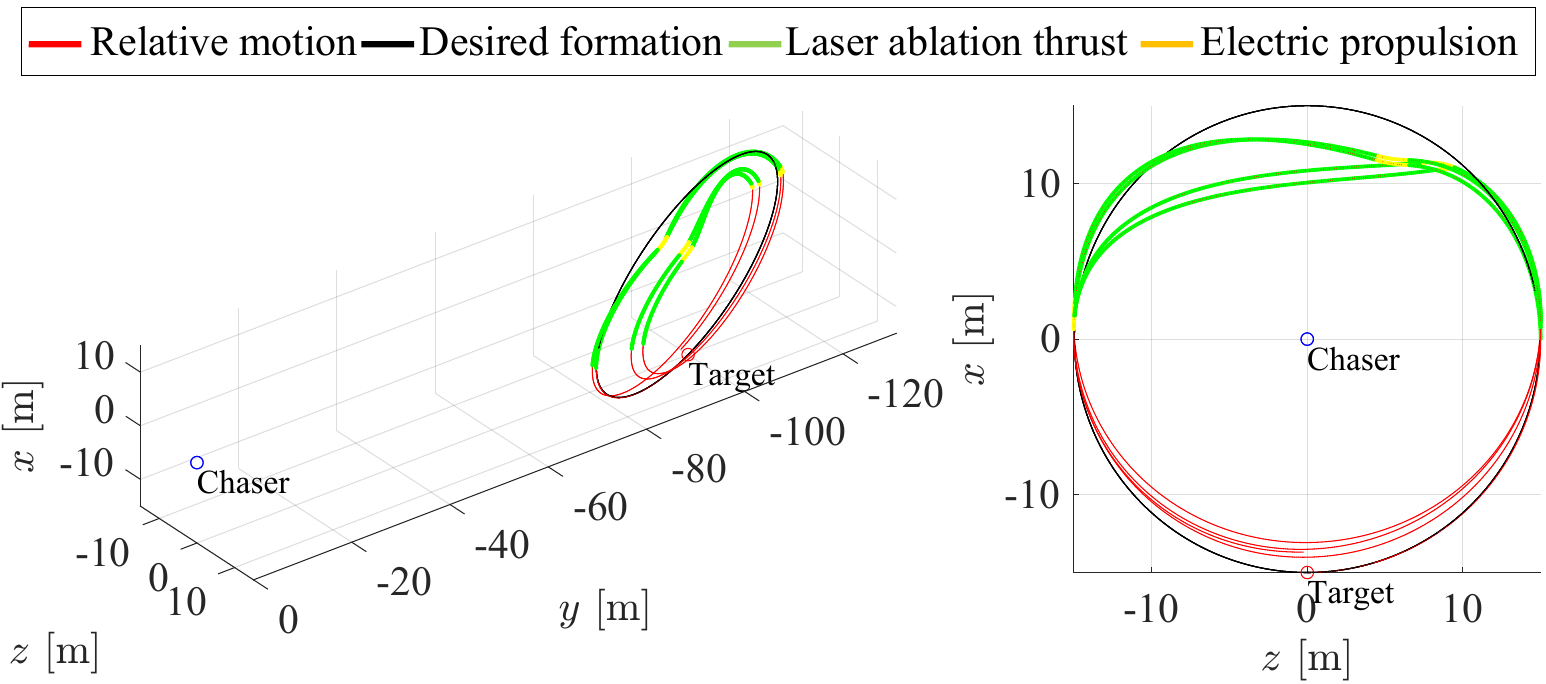}
\caption{Relative position for test case 1}
\label{fig_TC2_R_RM}
\end{figure*}
\begin{figure}[tb]
\centering
\includegraphics[scale = 0.3]{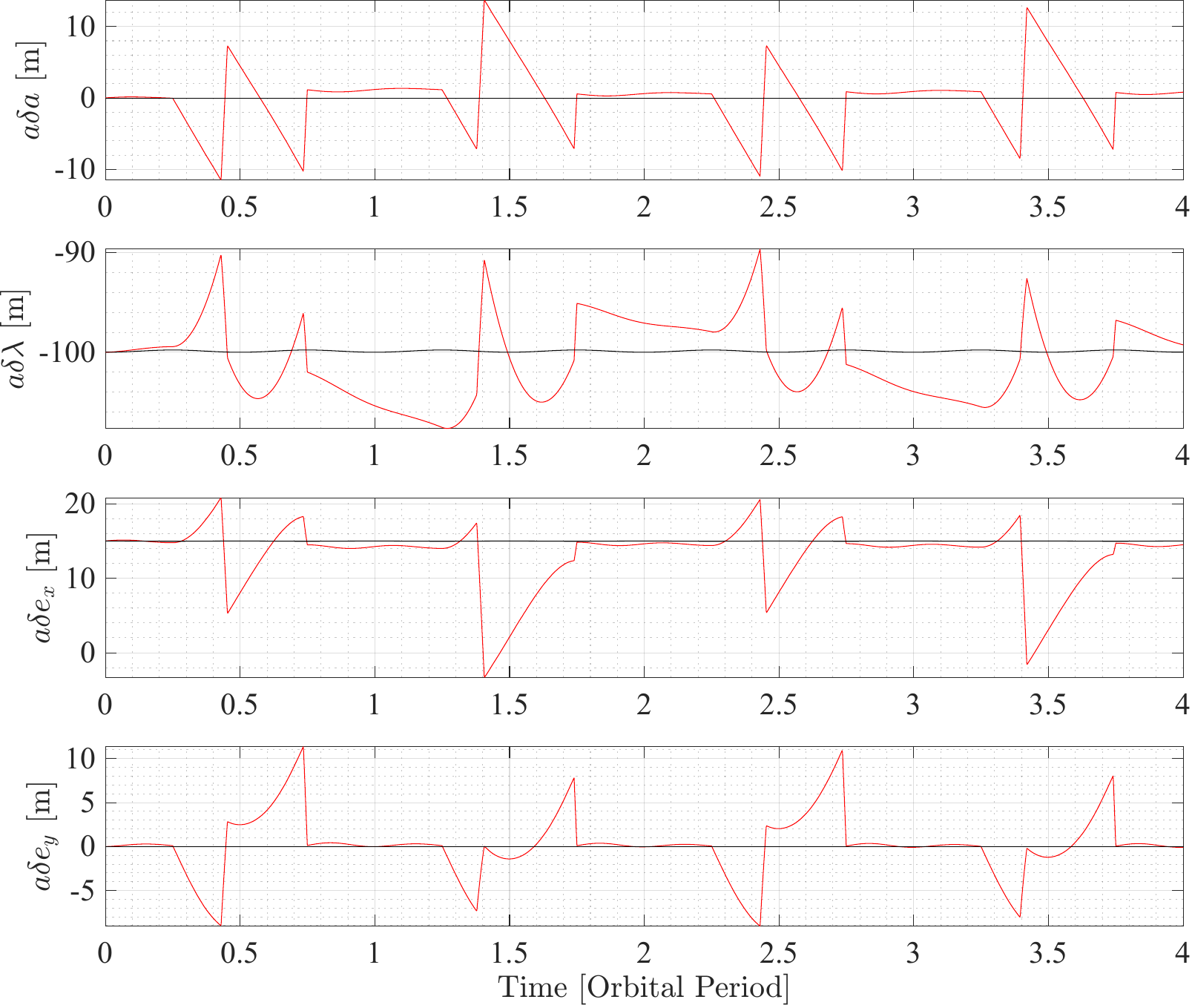}
\caption{ROE for test case 1}
\label{fig_TC2_R_aROE}
\end{figure}
\begin{figure}[tb]
\centering
\includegraphics[scale = 0.33]{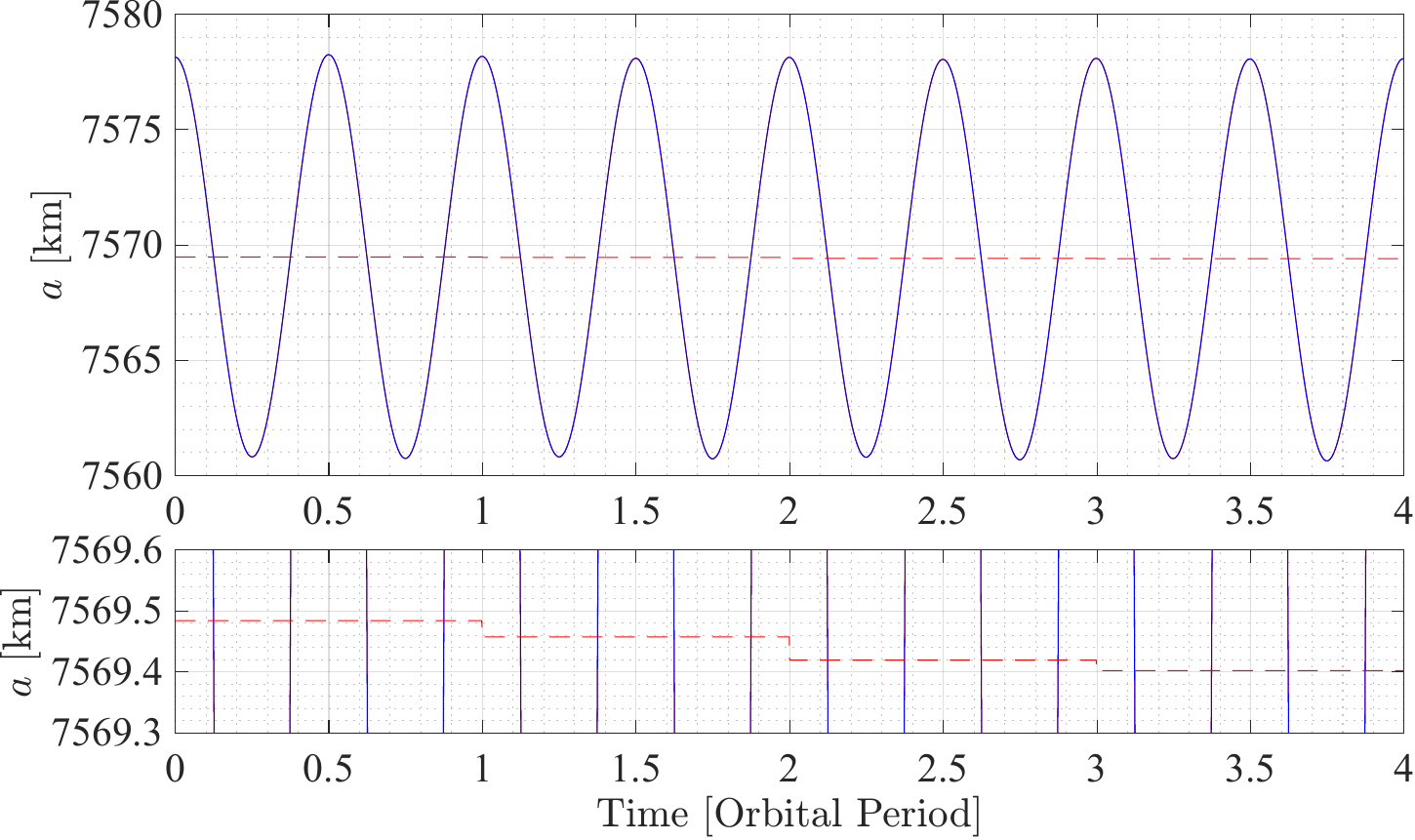}
\caption{Semi-major axis for test case 1 (top: full view, bottom: closeup)}
\label{fig_TC2_R_OE}
\end{figure}

%
Test case 1 applies the maneuver strategy 1 in Fig.~\ref{fig_MS1}. 
At the time to pass through perigee for each orbital revolution, the design parameters $t_{1, f}$, $t_{2, 0}$, $t_{2, f}$ and $\theta$ are determined by solving the proposed controller in Eqs.~\eqref{eq_control_des2} and~\eqref{eq_control_keeping_MS1}.  The initial time $t_{\rm ini}$ and the end time $t_m$ of the controller are simply computed as $t_{\rm ini} = (M-1)P$ and $t_m = MP$, where $P$ and $M$ are the orbital period and the number of orbital revolution, respectively.  In maneuver strategy 1, the first laser ablation start time $t_{1, 0}$ and the second electric propulsion end time $t_{3,0}$ are calculated as
\begin{align}
t_{1, 0} &= t_{\rm ini} + 0.25P \hspace{10pt} \rm{[s]}\\
t_{3,0} &= t_{\rm ini} + 0.75P \hspace{10pt} \rm{[s]}
\end{align}The proposed controller in Eqs.~\eqref{eq_control_des2} and~\eqref{eq_control_keeping_MS1} is solved by using Newton iteration process in this study.  The number of iterations depends on a solver's stopping criteria, including several tolerances.  In this simulation, the step tolerance and function tolerance is set to $1.0 \times 10^{-28}$ and $1.0 \times 10^{-14}$, respectively.  Newton iteration process is needed to determine the good initial guess for the iterative algorithm.  If the initial guess is too far from the true value, the iteration process may fail.  In this study, the initial guess of the design parameters is determined as follows.
\begin{align}
t_{1, f}^{(0)} &= t_{\rm ini} + 0.48P \hspace{10pt} \rm{[s]} \label{eq_InitialGuesse_MS1_1}\\
t_{2, 0}^{(0)} &= t_{\rm ini} + 0.5P \hspace{10pt} \rm{[s]} \label{eq_InitialGuesse_MS1_2}\\
t_{2, f}^{(0)} &= t_{\rm ini} + 0.73P \hspace{10pt} \rm{[s]} \label{eq_InitialGuesse_MS1_3}\\
\theta^{(0)} &= 180 \hspace{10pt} \rm{[deg]}
\end{align}These initial guesses are designed symmetrically with respect to the apogee in Fig.~\ref{fig_MS1}.  For this reason, the initial guess of the first electric propulsion end time $t_{2, 0}^{(0)}$ is set to 50\% of the orbital period as in Eq.~\eqref{eq_InitialGuesse_MS1_2}.
In addition, the initial guesses are designed so that the impulse generated by the laser ablation coincides with the impulse generated by electric propulsion.  These accelerations are calculated as follows from Table~\ref{table_con}.
\begin{align}
\frac{F_{\rm ab}/m_{t}}{F_{\rm el}/m_{c}} = \frac{4.80 \times 10^{-3}}{6.67 \times 10^{-2}} \sim \frac{4}{46}
\end{align}From these values of accelerations, the time duration of laser ablation and electric propulsion are set to 46\% and 4\% of the orbital period in the initial guesses of Eqs.~\eqref{eq_InitialGuesse_MS1_1}~and~\eqref{eq_InitialGuesse_MS1_3}.
In other words, the initial guess of the first electric propulsion interval and the second laser ablation interval are set to 2\% and 23\% of the orbital period, respectively.

%
Table~\ref{table_TC2} summarizes the design parameters obtained.   The number of iterations for solving the proposed controller was about 7 or 8.  Because the good initial guesses were selected, the iterative approach quickly converged.  Figure~\ref{fig_TC2_R_timing} describes the time history of the maneuver timings of the chaser (red) and the target (blue), including the thrust uncertainty.  The average time durations of the first laser ablation and electric propulsion over the four orbital revolution are about $17.3~\rm{min}$ and $2.7~\rm{min}$ (corresponding to 15.8\% and 2.5\% of the orbital period), respectively.  The average time durations of the second laser ablation and electric propulsion are about $33.3~\rm{min}$ and $1.3~\rm{min}$ (corresponding to 30.5\% and 1.2\% of the orbital period), respectively.  On the other hand, the $\theta$ varies significantly from $209.8~\rm{deg}$ to $145.2~\rm{deg}$ for each orbital revolution.  This electrical propulsion direction is adjusted to compensate for the disturbed relative orbit.

%
Figures~\ref{fig_TC2_R_RM} describes the relative position of the target.  
Adjusting the electrical propulsion (yellow) and laser ablation (green), the relative motion (red) periodically converges to the desired formation (black).  From this result, the formation keeping is confirmed for every orbital revolution.  

%
Figures~\ref{fig_TC2_R_aROE} describe the ROE of the target.  At the end of the orbital revolution, the relative mean longitude $\delta \lambda$ has the largest error in ROE components.  The reason for this error is an along-track drift motion of $\delta \lambda$ caused by the relative semi-major axis $\delta a$.  In this simulation, $\delta a$ tends to be larger because the actual ablation force $\hat{\bm{F}}_{\rm ab}$ (corresponding to the external force of the target) is smaller than the ideal value of $0.72~\rm{mN}$ in Table~\ref{table_con}.

%
The proposed controller aims to periodically return to the desired formation every orbital revolution. To compute the accuracy at the end of orbital revolution $M$, this study defined the error $\epsilon_M$ as follows.
\begin{align}
\epsilon_M = a(t_m) \| \delta \bm{\alpha}_{\rm in} (t_m) - \delta \bm{\alpha}_{\rm in,des} \| \hspace{10pt} t_m = MP
\end{align}where $\delta \bm{\alpha}_{\rm in} = [\delta a, \delta \lambda, \delta e_x, \delta e_y]^T$ is in-plane ROE components.  In this case, the error at each orbital revolution gradually decreases to $\epsilon_2 = 3.12~\rm{m}$, $\epsilon_3 = 4.79~\rm{m}$, and $\epsilon_4 = 1.72~\rm{m}$ from $\epsilon_1 = 5.64~\rm{m}$ at the first revolution.  The average error from $\epsilon_{30}$ to $\epsilon_{40}$ (corresponding to about 3 days) is $1.63~\rm{m}$. 
 Thus, these results verify the effectiveness of the proposed controller. 

%
Figure~\ref{fig_TC2_R_OE} describes the time history (blue) and the average (dotted red line) of the semi-major axes of the chaser and the target, which confirms their orbital decay.  In this case, the descent altitude over the four orbital revolutions and 3 days is $82.16~\rm{m}$ and $1131.5~\rm{m}$, respectively.
From these results, by periodically applying the proposed control law, both the chaser and the target can deorbit while maintaining their relative motion with respect to the target even under disturbances.

\subsection{Test case 2}

\begin{table}[tb]
    \caption{Design parameters for test case 2}
    \label{table_TC3}
    \centering
    \begin{tabular}{ccccc}
    \hline \hline
    Revolution, $M$ & $t_{1, 0} ~\rm{[s]}$ & $t_{2, f} ~\rm{[s]}$ & $\theta_1 ~\rm{[deg]}$& $\theta_2 ~\rm{[deg]}$ \\
    \hline
    1 & 1429.7 & 5131.6 & 326.0 & 214.0\\ 
    2 & 7957.7 & 11646 & 334.0 & 226.7\\ 
    3 & 14566 & 18256 & 325.8 & 213.8\\ 
    4 & 21092 & 24784 & 332.8 & 224.0\\ 
    \hline \hline
    \end{tabular}
\end{table}
\begin{figure}[tb]
\centering
\includegraphics[scale = 0.3]{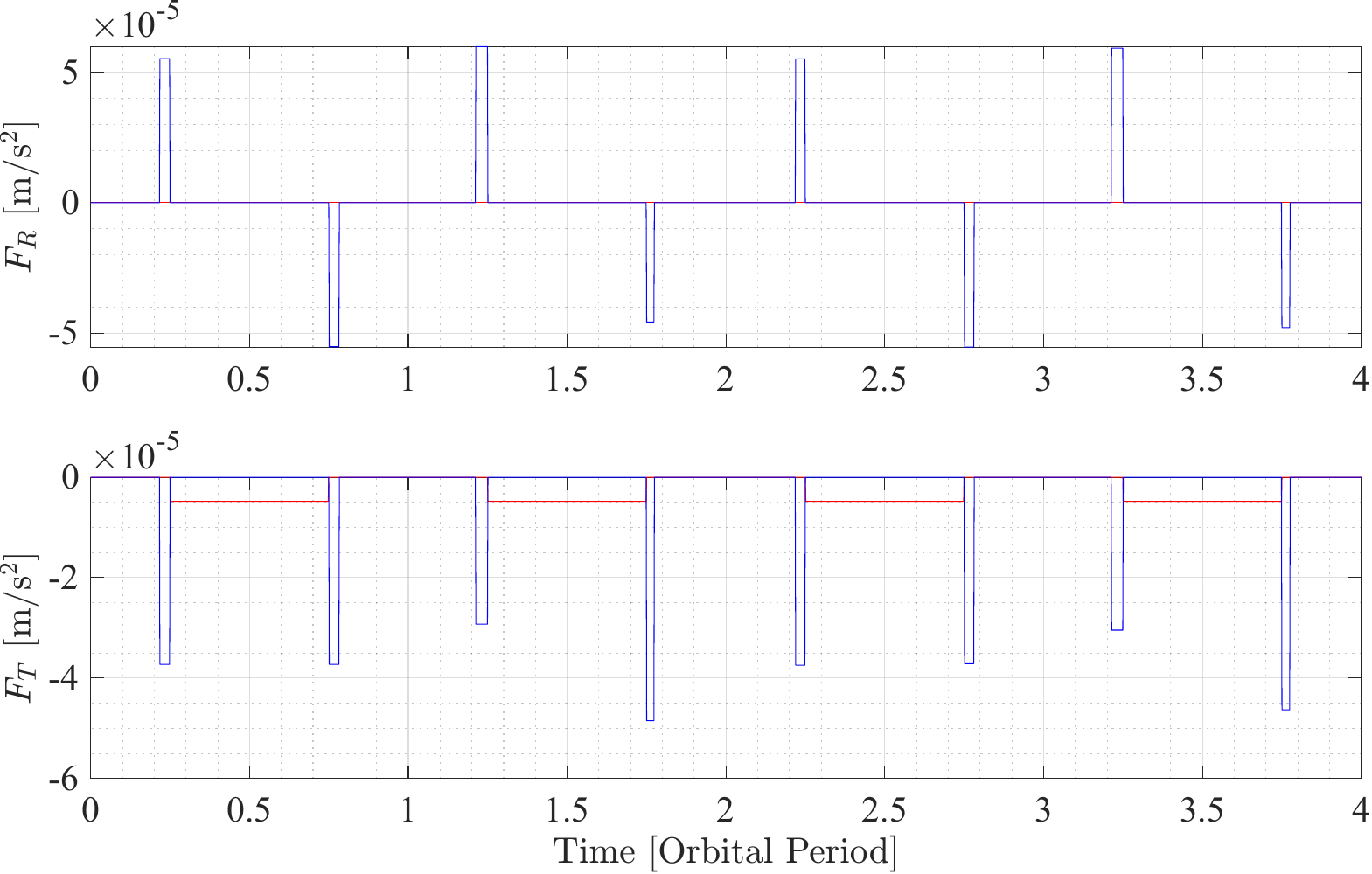}
\caption{Maneuver timing for test case 2}
\label{fig_TC3_R_timing}
\end{figure}
\begin{figure*}[tb]
\centering
\includegraphics[scale = 0.45]{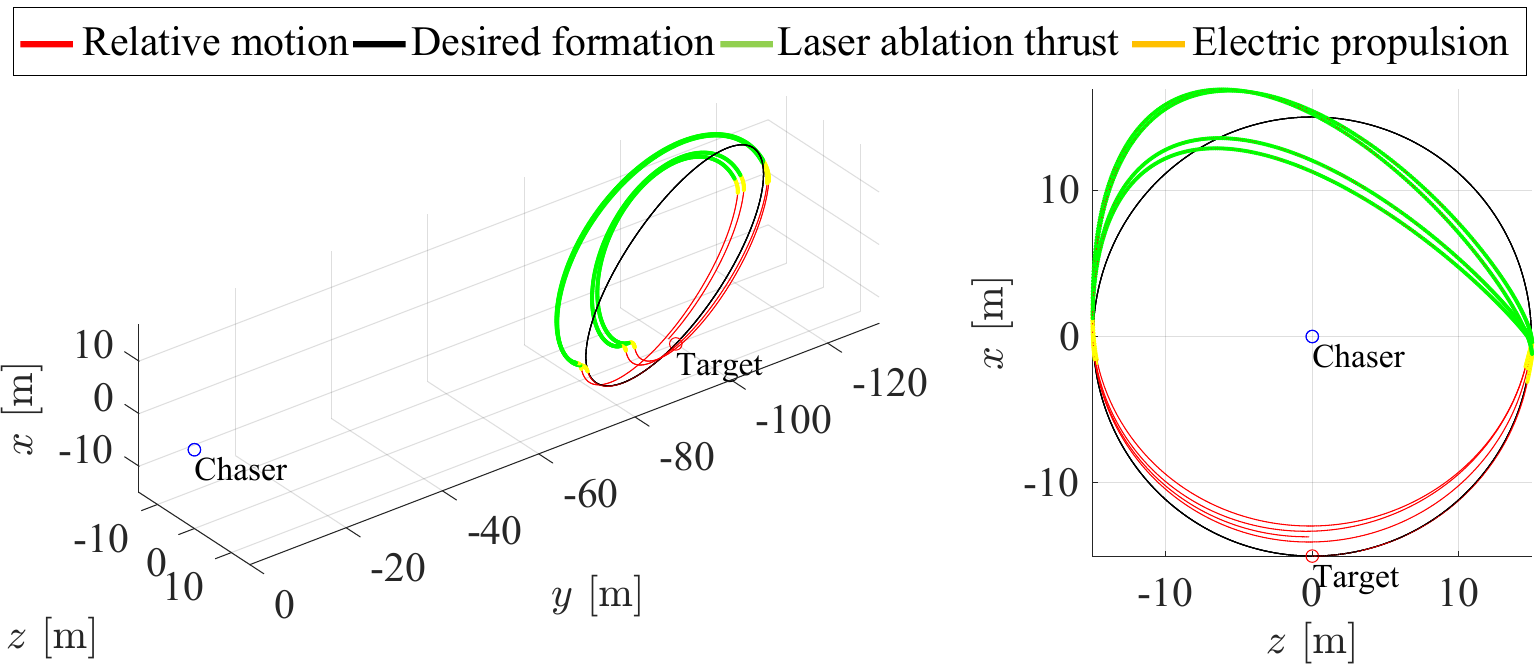}
\caption{Relative position for test case 2}
\label{fig_TC3_R_RM}
\end{figure*}
\begin{figure}[tb]
\centering
\includegraphics[scale = 0.3]{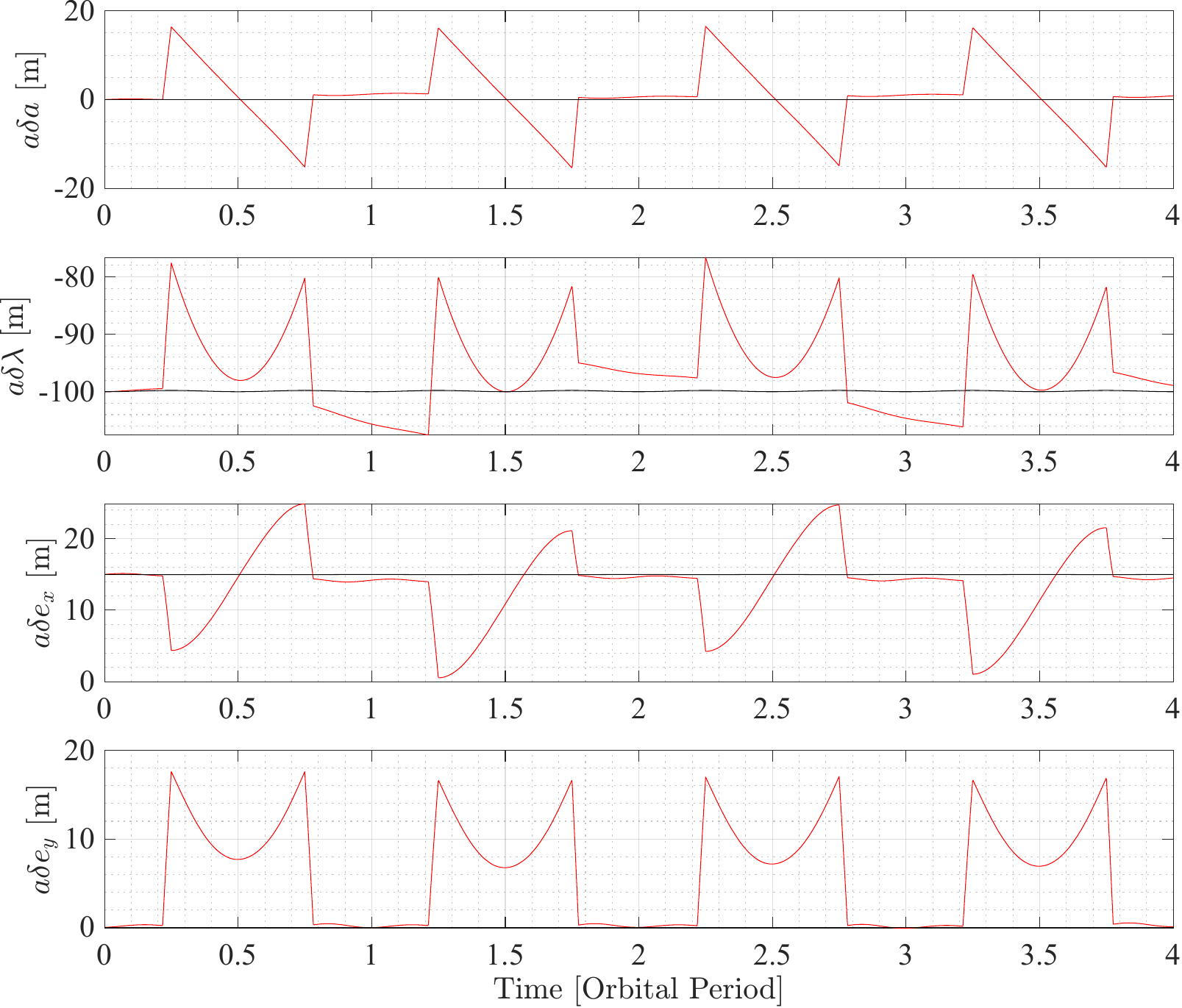}
\caption{ROE for test case 2}
\label{fig_TC3_R_aROE}
\end{figure}
\begin{figure}[tb]
\centering
\includegraphics[scale = 0.33]{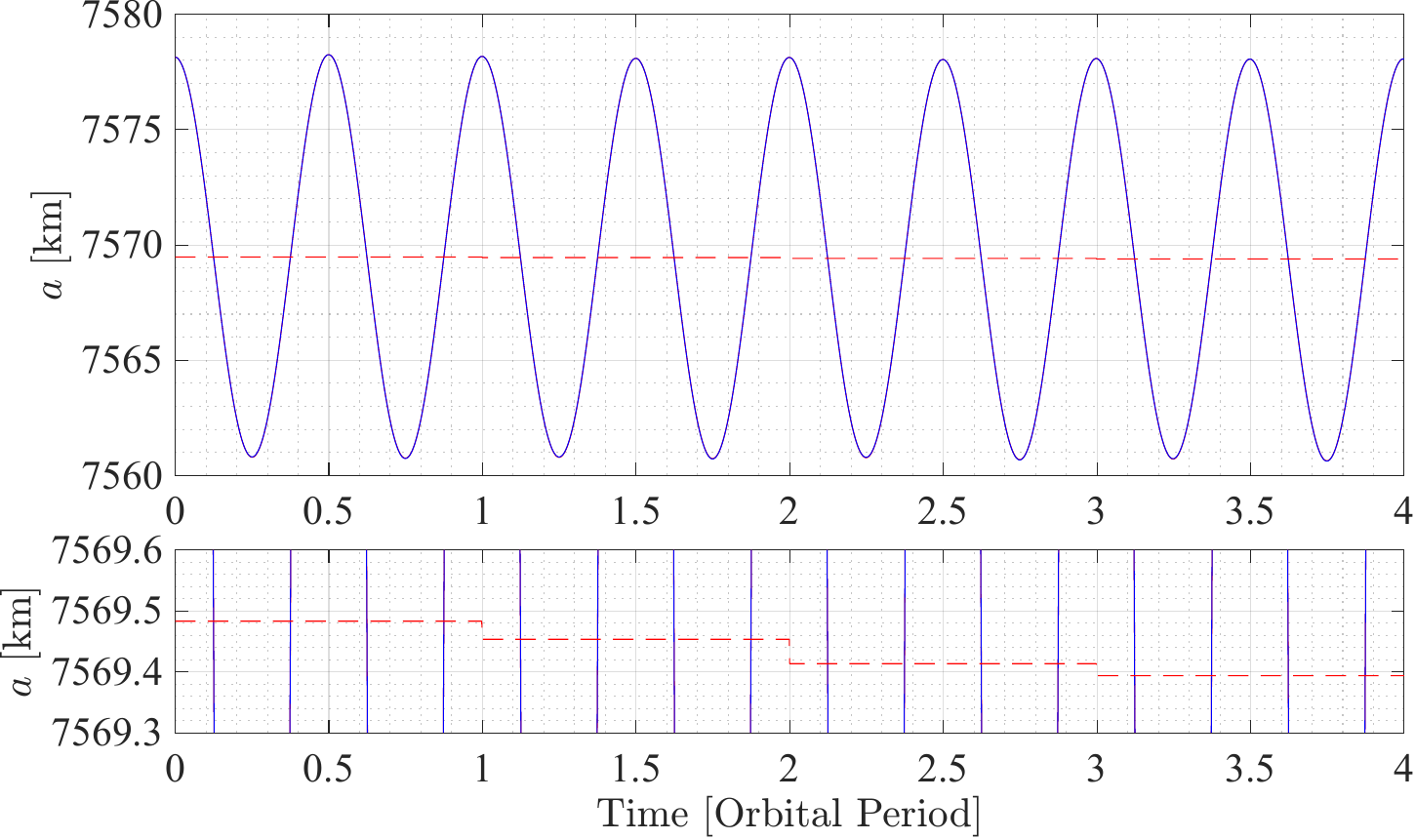}
\caption{Semi-major axis for test case 2 (top: full view, bottom: closeup)}
\label{fig_TC3_R_OE}
\end{figure}

%
Test case 2 applies the maneuver strategy 2 in Fig.~\ref{fig_MS2}.  
Similarly, at the time to pass through perigee for each orbital revolution, the design parameters $t_{1, 0}$, $t_{2, f}$, $\theta_1$ and $\theta_2$ are determined by solving Eqs.~\eqref{eq_control_des2} and~\eqref{eq_control_keeping_MS2} using Newton iteration process.  The laser ablation start time $t_{1, f}$ and the end time $t_{2,0}$ are calculated as
\begin{align}
t_{1, f} &= t_{\rm ini} + 0.25P \hspace{10pt} \rm{[s]}\\
t_{2,0} &= t_{\rm ini} + 0.75P \hspace{10pt} \rm{[s]}
\end{align}The solver's stopping criteria and the initial guess are selected by the same approach as Test case 1.  The initial guess of the design parameters is determined as follows.
\begin{align}
t_{1, 0}^{(0)} &= t_{\rm ini} + 0.23P \hspace{10pt} \rm{[s]} \label{eq_InitialGuesse_MS2_1}\\
t_{2, f}^{(0)} &= t_{\rm ini} + 0.77P \hspace{10pt} \rm{[s]} \label{eq_InitialGuesse_MS2_2}\\
\theta_1^{(0)} &= 270 \hspace{10pt} \rm{[deg]}\\
\theta_2^{(0)} &= 270 \hspace{10pt} \rm{[deg]}
\end{align}
%
Table~\ref{table_TC3} summarizes the results of design parameters.  The number of iterations for solving the proposed controller was about 6--8.  Figure~\ref{fig_TC3_R_timing} describes the time history of the maneuver timings of the chaser (red) and the target (blue) with uncertainty of thrust.  The maneuver timings and directions are adjusted to compensate for the disturbed relative orbit.  The average time durations of first and second electric propulsion over the four orbital revolution are about $3.7~\rm{min}$ and $3.1~\rm{min}$ (corresponding to 3.4\% and 2.8\% of the orbital period), respectively.

%
Figures~\ref{fig_TC3_R_RM} and~\ref{fig_TC3_R_aROE} describe the relative position and ROE of the target, which show the periodic return to the desired formation (black), which is similar to test case 1.  In Fig.~\ref{fig_TC3_R_aROE}, the large error of $\delta \lambda$ is also due to the same reason as test case 1.  The error at end of orbital revolution $M$ gradually decreases to $\epsilon_2 = 3.22~\rm{m}$, $\epsilon_3 = 4.79~\rm{m}$, and $\epsilon_4 = 1.46~\rm{m}$ from $\epsilon_1 = 5.82~\rm{m}$ at the first revolution.  The average error from $\epsilon_{30}$ to $\epsilon_{40}$ (corresponding to about 3 days) is $1.36~\rm{m}$.  Thus, these results verified the effectiveness of the proposed controller.

%
Figure~\ref{fig_TC3_R_OE} describes the time history of the semi-major axes of the chaser and target, which verifies their decay.  In this case, the descent altitude during the four orbital revolutions and 3 days is $89.4~\rm{m}$ and $1212.9~\rm{m}$, respectively.  This orbital decay is larger than test case 1, because the maneuver strategy 2 has longer laser ablation time than that for maneuver strategy 1.  
From these results, the control law proposed in this paper can be applied to various maneuver strategies.

%
Compared to Fig.~\ref{fig_TC2_R_RM}, the relative position (red) in Fig.~\ref{fig_TC3_R_RM} shows a large deformation relative to the desired formation (black).  This is due to the longer laser ablation time than the maneuver strategy 1 in Fig.~\ref{fig_MS1}.  This deformation reduces the minimum distance and could increase the collision probability with the target from the practical viewpoint.  From Tables~\ref{table_TC2} and~\ref{table_TC3}, the electrical propulsion directions $\theta$, $\theta_1$, and $\theta_2$ change significantly each time.  From the viewpoint of the long-term operation, the feasibility study of maneuver direction control is required.  Thus, this proposed control law is expected to contribute to the mission requirements definition for the actual laser ADR method.

%
In this study, the maneuver timings and directions are determined by solving four nonlinear equations in Eqs.~\eqref{eq_control_des2} and~\eqref{eq_control_keeping}.  In other words, four design parameters are selected in advance.  If the extra degrees of freedom are added increasing the number of design parameters, the optimization of the deorbiting time or fuel consumption is possible.  This consideration will be included in future works.

\subsection{Monte Carlo simulations}

\begin{figure}[tb]
\centering
\includegraphics[scale = 0.33]{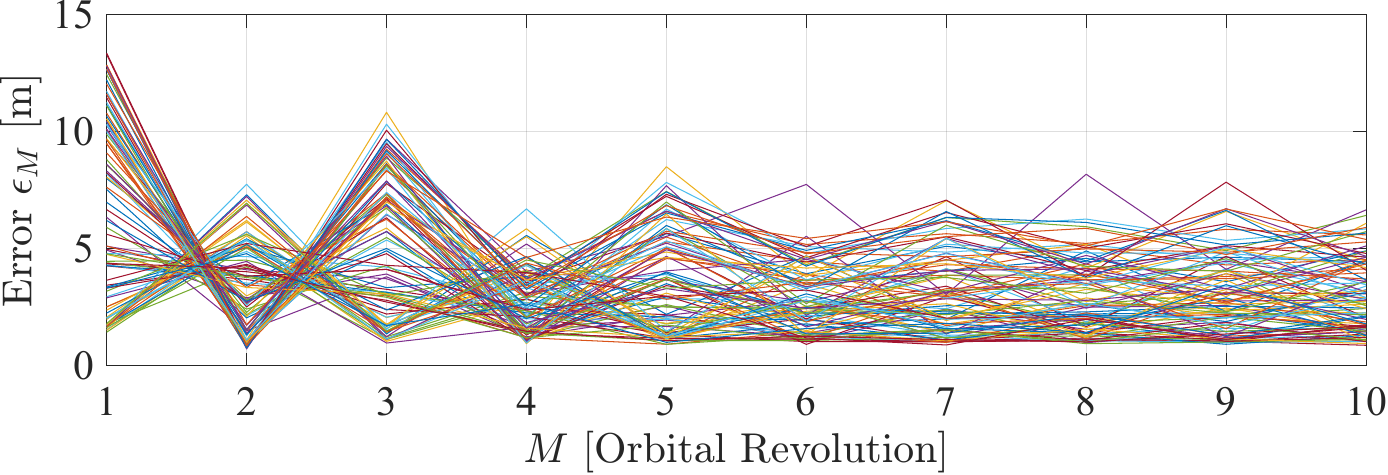}
\caption{Monte Carlo simulations}
\label{fig_MC}
\end{figure}

To verify the effectiveness of the proposed controller for arbitrary formations, 100 Monte Carlo runs are performed.
This Monte Carlo simulation considers maneuver strategy 2 because it has a longer laser ablation duration, which is practically useful for deorbiting.  
The desired formations are determined within the following range by sampling from uniform distribution.
\begin{align}
-1.5 \leq &a \delta a \leq 1.5 \hspace{10pt} \rm{[m]} \\
-150 \leq &a \delta \lambda \leq -100 \hspace{10pt} \rm{[m]} \\
-30 \leq &a \delta e_x \leq 30 \hspace{10pt} \rm{[m]} \\
-30 \leq &a \delta e_y \leq 30 \hspace{10pt} \rm{[m]} 
\end{align}Figure~\ref{fig_MC} describes the error $\epsilon_M$ $(M=1,\dots,10)$ at each orbital revolution.
The relative orbit errors are kept less than 8.2 m from $M=4$ to $M=10$, indicating the relative orbit is successfully controlled in all Monte Carlo runs.  Thus, this result verifies the effectiveness of the proposed controller for a wide variety of desired formation.


\section{Conclusions}

This paper dealt with the control law for simultaneous deorbiting of a chaser satellite and a target using laser ablation.
Although conventional formation flying missions assume that only a chaser maneuvers, this paper considers that both a chaser and a target have accelerations.  This paper derived the relative equations of motion between the chaser and the target in powered flight and their analytical solution.  Using the analytical solution, this paper proposed a control law for the simultaneous deorbit, which determines the timings and directions of the laser ablation and the electrical thrust so that the formation periodically returns to a desired formation.  The control law is flexible in terms of the desired formation and maneuver timings.  Therefore, this study examined two maneuver strategies to periodically return to the desired formation in one orbital revolution.

Numerical simulations verified the proposed control law under orbit perturbations and laser ablation uncertainty.  By periodically applying the proposed control law, both the chaser and the target can deorbit while maintaining their relative motion with respect to the target even under disturbances.  This flexibility of the control law will contribute to the design of mission operation and laser system specifications such as laser focal length, laser irradiating angle, and camera angle.  Future work will include optimizing the derived controller in terms of deorbiting time or fuel consumption, as well as the extension of the deorbit of multi-targets by one chaser.


\appendix
\section{}
\label{Appendix_nonlinear}
This section describes the nonlinear dynamics model of relative motion using the GVE.  Substituting Eqs.~\eqref{eq_gauss_a}--\eqref{eq_gauss_Omega} into Eq.~\eqref{eq_dif_roe} yields the following dynamics of relative motion.
\begin{align}
\delta \dot{\bm{\alpha}} (t) &= \bm{\zeta}(\delta\bm{\alpha},\bm{a}) = \begin{bmatrix} 0\\ n_t - n\\ 0\\ 0\\ 0\\ 0 \end{bmatrix} + B' \bm{a} 
\end{align}The elements of the matrix $B'$ are
\begin{align}
b'_{1 1} =& \frac{2\,e_{t}\,s_{f_{t}}}{a\,n_{t}\,{\sqrt{1-{e_{t}}^2}}}\\
b'_{1 2} =& \frac{2\,\left(e_{t}\,c_{f_{t}}+1\right)}{a\,n_{t}\,{\sqrt{1-{e_{t}}^2}}} \\
b'_{1 3} =& 0\\
b'_{1 4} =& -\frac{2\,a_{t}\,e\,s_{f}}{a^2\,n\,{\sqrt{1-e^2}}}\\
b'_{1 5} =& -\frac{2\,a_{t}\,\left(e\,c_{f}+1\right)}{a^2\,n\,{\sqrt{1-e^2}}}\\
b'_{1 6} =& 0\\
b'_{2 1} =& -\frac{a_{t}\,c_{f_{t}}\,\left({e_{t}}^2-1\right)-\frac{2\,a_{t}\,e_{t}\,\left({e_{t}}^2-1\right)}{e_{t}\,c_{f_{t}}+1}}{{a_{t}}^2\,e_{t}\,n_{t}}-\frac{{\sqrt{1-{e_{t}}^2}}\,c_{f_{t}}}{a_{t}\,e_{t}\,n_{t}}\\
b'_{2 2} =& \frac{s_{f_{t}}\,\left(e_{t}\,c_{f_{t}}+2\right)\,\left({\sqrt{1-{e_{t}}^2}}+{e_{t}}^2-1\right)}{a_{t}\,e_{t}\,n_{t}\,\left(e_{t}\,c_{f_{t}}+1\right)}\\ 
b'_{2 3} =& \frac{{\sqrt{1-{e_{t}}^2}}\,s_{\theta _{t}}\,\left(c_{i}-c_{i_{t}}\right)}{a_{t}\,n_{t}\,s_{i_{t}}\,\left(e_{t}\,c_{f_{t}}+1\right)}\\
b'_{2 4} =& \frac{a\,c_{f}\,\left(e^2-1\right)-\frac{2\,a\,e\,\left(e^2-1\right)}{e\,c_{f}+1}}{a^2\,e\,n}+\frac{{\sqrt{1-e^2}}\,c_{f}}{a\,e\,n}\\
b'_{2 5} =& -\frac{s_{f}\,\left(e\,c_{f}+2\right)\,\left({\sqrt{1-e^2}}+e^2-1\right)}{a\,e\,n\,\left(e\,c_{f}+1\right)}\\ 
b'_{2 6} =& \frac{{\sqrt{1-e^2}}\,c_{i}\,c_{\theta}\,\left(\Omega -\Omega _{t}\right)}{a\,n\,\left(e\,c_{f}+1\right)}\\
b'_{3 1} =& \frac{{\sqrt{1-{e_{t}}^2}}\,s_{\theta_t}}{a_{t}\,n_{t}}\\
b'_{3 2} =& \frac{{\sqrt{1-{e_{t}}^2}}\,\left(2\,c_{\theta_t}+\frac{3\,e_{t}\,c_{\omega _{t}}}{2}+\frac{e_{t}\,c_{2f_{t}+\omega _{t}}}{2}\right)}{a_{t}\,n_{t}\,\left(e_{t}\,c_{f_{t}}+1\right)}\\
b'_{3 3} =& \frac{e_{t}\,{\sqrt{1-{e_{t}}^2}}\,\mathrm{cot}\left(i_{t}\right)\,s_{\omega _{t}}\,s_{\theta _{t}}}{a_{t}\,n_{t}\,\left(e_{t}\,c_{f_{t}}+1\right)} \\
b'_{3 4} =& -\frac{{\sqrt{1-e^2}}\,s_{\theta }}{a\,n}
\end{align}\begin{align}
b'_{3 5} =& -\frac{{\sqrt{1-e^2}}\,\left(2\,c_{\theta }+\frac{3\,e\,c_{\omega }}{2}+\frac{e\,c_{2f+\omega}}{2}\right)}{a\,n\,\left(e\,c_{f}+1\right)}\\
b'_{3 6} =& -\frac{e\,{\sqrt{1-e^2}}\,\mathrm{cot}\left(i\right)\,s_{\omega }\,s_{\theta}}{a\,n\,\left(e\,c_{f}+1\right)} \\
b'_{4 1} =& -\frac{{\sqrt{1-{e_{t}}^2}}\,c_{\theta_t}}{a_{t}\,n_{t}}\\
b'_{4 2} =& \frac{{\sqrt{1-{e_{t}}^2}}\,\left(2\,s_{\theta_t}+\frac{3\,e_{t}\,s_{\omega _{t}}}{2}+\frac{e_{t}\,s_{2f_{t}+\omega _{t}}}{2}\right)}{a_{t}\,n_{t}\,\left(e_{t}\,c_{f_{t}}+1\right)}\\ 
b'_{4 3} =& -\frac{e_{t}\,{\sqrt{1-{e_{t}}^2}}\,\mathrm{cot}\left(i_{t}\right)\,c_{\omega _{t}}\,s_{\theta _{t}}}{a_{t}\,n_{t}\,\left(e_{t}\,c_{f_{t}}+1\right)}\\
b'_{4 4} =& \frac{{\sqrt{1-e^2}}\,c_{\theta }}{a\,n}\\ 
b'_{4 5} =& -\frac{{\sqrt{1-e^2}}\,\left(2\,s_{\theta }+\frac{3\,e\,s_{\omega }}{2}+\frac{e\,s_{2f+\omega}}{2}\right)}{a\,n\,\left(e\,c_{f}+1\right)}\\
b'_{4 6} =& \frac{e\,{\sqrt{1-e^2}}\,\mathrm{cot}\left(i\right)\,c_{\omega }\,s_{\theta}}{a\,n\,\left(e\,c_{f}+1\right)}\\
b'_{5 1} =& b'_{5 2} = 0\\ 
b'_{5 3} =& \frac{{\sqrt{1-{e_{t}}^2}}\,\cos\left(\theta _{t}\right)}{a_{t}\,n_{t}\,\left(e_{t}\,c_{f_{t}}+1\right)}\\
b'_{5 4} =& b'_{5 5} = 0\\
b'_{5 6} =& -\frac{{\sqrt{1-e^2}}\,c_{\theta}}{a\,n\,\left(e\,c_{f}+1\right)}\\
b'_{6 1} =& b'_{6 2} = 0\\ 
b'_{6 3} =& \frac{{\sqrt{1-{e_{t}}^2}}\,s_{i }\,s_{\theta _{t}}}{a_{t}\,n_{t}\,s_{i_{t}}\,\left(e_{t}\,c_{f_{t}}+1\right)}\\
b'_{6 4} =& b'_{6 5} = 0\\ 
b'_{6 6} =& -\frac{{\sqrt{1-e^2}}\,\left(s_{\theta}+\Omega \,s_{i }\,c_{\theta}-\Omega _{t}\,s_{i }\,c_{\theta}\right)}{a\,n\,\left(e\,c_{f}+1\right)}
\end{align}where $f$ and $\theta$ represent true anomaly and true argument of latitude, respectively.  Performing a first-order Taylor expansion of this nonlinear function around the chaser orbit ($\delta\bm{\alpha}=0$) yield the following linear dynamics Eq.~\eqref{eq_form_dif}.

\section{}
\label{Appendix_transformation}
This section describes detailed transformation from Eq.~\eqref{eq_form0} to Eq.~\eqref{eq_form}.
\begin{align}
\int_{t_0}^{t} \Phi(\tau, t_0, \bm{a}_{c,{\rm in}}) B\bm{a} d \tau = \left[ \int_{t_0}^{t} \Phi(\tau, t_0, \bm{a}_{c,{\rm in}}) B d \tau \right] \bm{a}
\end{align}Neglecting the chaser maneuver of out-plane direction 
\begin{align}
\left[ \int_{t_0}^{t} \Phi(\tau, t_0, \bm{a}_{c,{\rm in}}) B d \tau \right] \bm{a} = \Gamma \begin{bmatrix} F_{Rt}/m_t\\ F_{Tt}/m_t\\ F_{Nt}/m_t\\ F_{R}/m_c\\ F_{T}/m_c \end{bmatrix}
\end{align}
If the chief maneuvers are not impulsive thrust but low thrust, this paper gets
\begin{equation}
\label{eq_approximation}
\frac{1 +J(t-t_0)}{W_{1} +2W_2(t-t_0)} \approx \frac{1}{W_1}
\end{equation}The elements of the matrix $\Gamma$ can be expressed as follows using Eqs.~\eqref{eq_a_form1}, \eqref{eq_n_form1}, \eqref{eq_u_form1} and the approximation Eq.~\eqref{eq_approximation}

\begin{align}
\gamma_{2 1} =& - \int_{t_0}^{t} \frac{2}{a(t)n(t)} d \tau\\
=& - \int_{t_0}^{t} \frac{2}{a_0 n_0}\left(1 +J(t-t_0)\right) d \tau\\
=& -\frac{(J(t-t_0)+2)(t-t_0)}{a_0 n_0}\\
=& -\frac{\Psi_{21}}{a_0 n_0 W_1}\\
\gamma_{3 1} =& \int_{t_0}^{t} \frac{s_u}{a(t)n(t)} d \tau\\
=& \frac{1}{a_0 n_0} \int_{u_0}^{u} \left(1 +J(t-t_0)\right) s_\upsilon \frac{d \upsilon}{W_{1} +2W_2(t-t_0)}\\
=& \frac{1}{a_0 n_0} \int_{u_0}^{u} \frac{1 +J(t-t_0)}{W_{1} +2W_2(t-t_0)} s_\upsilon d \upsilon\\
\approx & \frac{2}{a_0 n_0} \int_{u_0}^{u} \frac{1}{W_{1}} s_\upsilon d \upsilon\\
=& \frac{1}{a_0 n_0 W_1}(c_{u0} - c_u)
\end{align}\begin{align}
\gamma_{4 1} =& - \int_{t_0}^{t} \frac{c_u}{a(t)n(t)} d \tau\\
=& - \frac{1}{a_0 n_0} \int_{u_0}^{u} \left(1 +J(t-t_0)\right) c_\upsilon \frac{d \upsilon}{W_{1} +2W_2(t-t_0)}\\
=& - \frac{1}{a_0 n_0} \int_{u_0}^{u} \frac{1 +J(t-t_0)}{W_{1} +2W_2(t-t_0)} c_\upsilon d \upsilon\\
\approx & - \frac{1}{a_0 n_0} \int_{u_0}^{u} \frac{1}{W_{1}} c_\upsilon d \upsilon\\
=& \frac{1}{a_0 n_0 W_1}(s_{u0} - s_u)\\
\gamma_{2 2} =& \int_{t_0}^{t} \frac{2}{a(t)n(t)}\Phi_{21} d \tau\\
=& \int_{t_0}^{t} \frac{2}{a_0 n_0}\left(1 +J(t-t_0)\right)\Phi_{21} d \tau\\
=& \frac{(9J^2(t-t_0)^2+4J(t-t_0)-12)(t-t_0)^2}{8 a_0}\\
=& \frac{\Psi_{22}}{a_0 n_0 W_1}
\end{align}Performing similar transformation for each component
\begin{align}
\gamma_{1 1} =& \gamma_{5 1} = \gamma_{6 1} = 0\\
\gamma_{1 2} =& \frac{\Psi_{21}}{a_0 n_0 W_1}\\
\gamma_{3 2} =& \frac{2}{a_0 n_0 W_1}(s_u - s_{u0})\\
\gamma_{4 2} =& - \frac{2}{a_0 n_0 W_1}(c_u - c_{u0})\\
\gamma_{5 2} =& \gamma_{6 2} = \gamma_{1 3} = \gamma_{2 3} = \gamma_{3 3} = \gamma_{4 3} = 0\\
\gamma_{5 3} =& \frac{1}{a_0 n_0 W_1}(s_{u0} - s_u)\\
\gamma_{6 3} =& \frac{1}{a_0 n_0 W_1}(c_{u0} - c_u) \\
\gamma_{1 4} =& 0 \\
\gamma_{2 4} =& \frac{\Psi_{21}}{a_0 n_0 W_1}\\
\gamma_{3 4} =& \frac{1}{a_0 n_0 W_1}(c_u - c_{u0})\\
\gamma_{4 4} =& \frac{2}{a_0 n_0 W_1}(s_u - s_{u0})\\
\gamma_{5 4} =& \gamma_{6 4} = 0\\
\gamma_{1 5} =& -\frac{\Psi_{21}}{a_0 n_0 W_1}\\
\gamma_{2 5} =& -\frac{\Psi_{22}}{a_0 n_0 W_1}\\
\gamma_{3 5} =& -\frac{2}{a_0 n_0 W_1}(s_u - s_{u0})\\
\gamma_{4 5} =& \frac{2}{a_0 n_0 W_1}(c_u - c_{u0})\\ 
\gamma_{5 5} =& \gamma_{6 5} = 0
\end{align}Decomposing this matrix $\Gamma$
\begin{align}
\Gamma \begin{bmatrix} F_{Rt}/m_t\\ F_{Tt}/m_t\\ F_{Nt}/m_t\\ F_{R}/m_c\\ F_{T}/m_c \end{bmatrix} &= \Psi_c (t, t_0, \bm{a}_{c,{\rm in}}) \begin{bmatrix} F_{R}/m_c\\ F_{T}/m_c \end{bmatrix} + \Psi_t (t, t_0, \bm{a}_{c,{\rm in}}) \begin{bmatrix} F_{Rt}/m_t\\ F_{Tt}/m_t\\ F_{Nt}/m_t \end{bmatrix} \\
&= \Psi_c (t, t_0, \bm{a}_{c,{\rm in}}) \bm{a}_{c,{\rm in}} + \Psi_t (t, t_0, \bm{a}_{c,{\rm in}}) \bm{a}_t
\end{align}


\bibliographystyle{jasr-model5-names} 
\biboptions{authoryear} 
\bibliography{isobe.bib}

\end{document}